\documentclass{acmart}

\AtBeginDocument{%
  \providecommand\BibTeX{{%
    \normalfont B\kern-0.5em{\scshape i\kern-0.25em b}\kern-0.8em\TeX}}}

\setcopyright{acmlicensed}
\acmJournal{TKDD}
\acmYear{2024} 
\acmVolume{1} 
\acmNumber{1} 
\acmArticle{1} 
\acmMonth{1}
\acmDOI{10.1145/3694980}

\acmPrice{15.00}
\acmISBN{978-1-4503-XXXX-X/18/06}

\newcommand{\new}[1]{#1}
\newcommand{\old}[1]{} 
\usepackage{subcaption}
\usepackage{multirow}
\usepackage{tikz}
\usepackage[edges]{forest}
\usepackage{fancyvrb}
\usepackage{soul}
\usepackage[flushleft]{threeparttable}
\usepackage{tablefootnote}
\usepackage{pifont}
\usepackage{pdfrender}
\usepackage[normalem]{ulem} % for strike through support
\newcommand{\dashcheckmark}{
    \textpdfrender{
        TextRenderingMode=Stroke,
        LineWidth=1.0pt,
        LineDashPattern=[1 2]0,
    }{\checkmark}
}

\newcommand{\cmark}{\ding{51}}%
\usepackage{setspace} 
\newcommand{\ie}{{i.e.}}
\newcommand{\eg}{{e.g.}}

\begin{document}

\title{A Survey on the Role of Crowds in Combating Online Misinformation: Annotators, Evaluators, and Creators}

\author{Bing He}
\affiliation{%
  \institution{Georgia Institute of Technology}
  \country{USA}}
\email{bhe46@gatech.edu}

\author{Yibo Hu}
\affiliation{%
  \institution{Georgia Institute of Technology}
  \country{USA}}
\email{yibo.hu@gatech.edu}

\author{Yeon-Chang Lee}
\affiliation{%
  \institution{Ulsan National Institute of Science and Technology (UNIST)}
  \country{Korea}}
\email{yeonchang@unist.ac.kr}

\author{Soyoung Oh}
\affiliation{%
  \institution{Saarland University}
  \country{Germany}}
\email{soyoung@lst.uni-saarland.de}

\author{Gaurav Verma}
\affiliation{%
  \institution{Georgia Institute of Technology}
  \country{USA}}
\email{gverma@gatech.edu}

\author{Srijan Kumar}
\affiliation{%
  \institution{Georgia Institute of Technology}
  \country{USA}}
\email{srijan@gatech.edu}


\begin{abstract}

Online misinformation poses a global risk with significant real-world consequences. To combat misinformation, current research relies on professionals like journalists and fact-checkers for annotating and debunking false information, while also developing automated machine learning methods for detecting misinformation. Complementary to these approaches, recent research has increasingly concentrated on utilizing the power of ordinary social media users, a.k.a. ``the crowd'', who act as eyes-on-the-ground proactively questioning and countering misinformation.  Notably, recent studies show that 96\% of counter-misinformation responses originate from them. Acknowledging their prominent role, we present the first systematic and comprehensive survey of research papers that actively leverage the crowds to combat misinformation.

In this survey, we first identify 88 papers related to crowd-based efforts\footnote{\url{https://github.com/claws-lab/awesome-crowd-combat-misinformation}}, following a meticulous annotation process adhering to the PRISMA framework (preferred reporting items for systematic reviews and meta-analyses). 
We then present key statistics related to misinformation, counter-misinformation, and crowd input in different formats and topics.
Upon holistic analysis of the papers, we introduce a novel taxonomy of the roles played by the crowds in combating misinformation:
(i) \textit{crowds as annotators} who actively identify misinformation;
(ii) \textit{crowds as evaluators} who assess counter-misinformation effectiveness;
(iii) \textit{crowds as creators} who create counter-misinformation.
This taxonomy explores the crowd's capabilities in misinformation detection, identifies the prerequisites for effective counter-misinformation, and analyzes crowd-generated counter-misinformation. 
In each assigned role, we conduct a detailed analysis to categorize the specific utilization of the crowd. Particularly, we delve into (i) distinguishing individual, collaborative, and machine-assisted labeling for annotators; (ii)  analyzing the effectiveness of counter-misinformation through surveys, interviews, and in-lab experiments for evaluators; and (iii) characterizing creation patterns and creator profiles for creators. 
Finally, we conclude this survey by outlining potential avenues for future research in this field.

\end{abstract}

\begin{CCSXML}
<ccs2012>
   <concept>
       <concept_id>10002951.10003227.10003351</concept_id>
       <concept_desc>Information systems~Data mining</concept_desc>
       <concept_significance>500</concept_significance>
       </concept>
   <concept>
       <concept_id>10010147.10010257</concept_id>
       <concept_desc>Computing methodologies~Machine learning</concept_desc>
       <concept_significance>500</concept_significance>
       </concept>
   <concept>
       <concept_id>10002951.10003260.10003282.10003292</concept_id>
       <concept_desc>Information systems~Social networks</concept_desc>
       <concept_significance>500</concept_significance>
       </concept>
 </ccs2012>
\end{CCSXML}

\ccsdesc[500]{Information systems~Data mining}
\ccsdesc[500]{Information systems~Social networks}

\keywords{Misinformation, Combat misinformation, Crowd, Survey, Counter-misinformation}

\maketitle

\section{Introduction}

Most individuals today rely on social media platforms as their primary source of news and information \cite{walker2021news}. 
However, such platforms contain a plethora of unreliable information, including misinformation, which unfortunately spreads more rapidly and widely than truth \cite{vosoughi2018spread}. 
Online misinformation harms individuals and society at multiple levels. 
At the micro-level, misinformation harms well-being~\cite{verma2022examining}, increases polarization~\cite{stewart2018examining}, and leads to online harassment and violent attacks on individuals and communities~\cite{Chua2017}. 
At the macro-level, misinformation questions democratic processes and elections~\cite{silverman2016analysis},
impacts science and global public health~\cite{Memon2020, dai2022effects, xue2022covid}. 
For instance, misinformation about the COVID-19 vaccine (e.g., the vaccine causes infertility) has reduced vaccine uptake and prolonged the pandemic \cite{Memon2020, dai2022effects, xue2022covid}; misinformation about elections can \new{undermine}  trust in democratic processes and institutions \cite{Cohen2020}.
Therefore, it is crucial to curb the spread of online misinformation and to counter misinformation \cite{Sun2020, Zhao2016, vosoughi2018spread}.

Motivated by this, research has focused on detecting misinformation by utilizing different approaches, including \textbf{automated machine learning (ML) solutions}~\cite{Sharma2019, guo2019future, islam2020deep} and the use of \textbf{professional fact-checkers}~\cite{micallef2022true, porter2021global, markowitz2023cross}.
Notably, there has been a growing interest in research on developing ML models~\cite{Sharma2019} based on post content, poster attributes, social network, temporal aspects,  and propagation features~\cite{shu2020fakenewsnet}.
These models have been deployed across widely-used web and social media platforms (\eg, Twitter (\new{
Twitter was renamed as ``X'' in July 2023. We continue to refer to the platform as ``Twitter'' for illustration in this survey.}), Facebook, and YouTube).  
In the meantime, ML solutions rely on ground truth labels of misinformation for their training and validation, whereas professional fact-checkers typically label the misinformation with fact-check labels.
These professionals also write fact-checking articles to explain their reasoning for the label determinations. 
For example, \textit{Snopes.com} provides fact-check labels that range from ``true'' and ``mostly true'' to ``mostly false'' and ``false'', accompanied by corresponding explanations.

Despite these solutions, the ``infodemic'', or the epidemic of misinformation~\cite{Borah2021, chen2022let, xue2022covid}, continues to grow at an alarming rate.
One contributing factor is that automated ML models respond slowly to changes in the information ecosystem, rely on fact-check labels provided by professional fact-checkers, and are vulnerable to manipulation by adversaries \cite{Allen2021, he2021petgen, roitero2021can}.
On the other hand, professional fact-checkers face \new{constraints}  in terms of the limited number of fact-checkers and the significant time required for label generation; moreover, their fact-checks tend to address only a small number of viral claims~\cite{Allen2021, Pennycook2019, micallef2022true}.
Importantly, both approaches only detect misinformation but do not actively engage with misinformation spreaders.
In this context, it is worth noting that there has been a noticeable absence of discussion on structured methods for countering misinformation once it is identified.

\subsection{Motivation}\label{sec:motivation}

To address the drawbacks of these two approaches, leveraging \textbf{crowds} offers a promising solution in a scalable and proactive manner \cite{Allen2021, ma2023characterizing}.
In this study, we focus on the ``crowd'' --- defined as \textit{ordinary users of social media platforms} (\ie, not fact-checkers, journalists, or organizations).
They serve as eyes on the ground who proactively question and counter misinformation, including emerging misinformation~\cite{Bode2018, he2023reinforcement}.
Literature has shown that crowd-based ``social correction'' is effective and works well across topics~\cite{Bode2018, ma2023characterizing}. 
\new{
Crowds combat misinformation by contributing significantly to diverse tasks} 
, including identifying misinformation, assessing counter-misinformation effectiveness, and creating counter-misinformation. 
In addition, crowds also have the key benefit of being ``\textbf{cost-effective} '' \new{by countering online misinformation voluntarily for collective efforts}~\cite{mujumdar2021hawkeye, allen2022birds},  
\new{especially when} compared to the expensive and time-consuming recruitment of professional fact-checkers. 
Therefore, there has been a surge of research efforts to develop \ul{crowd-based methodologies to annotate misinformation, evaluate counter-misinformation effectiveness, and characterize counter-misinformation creation} in recent years \cite{Bhuiyan2020, chen2021citizens, zhang2022investigation, wang2022factors, wang2021evaluating, Allen2021, mujumdar2021hawkeye, Shabani2018}. 
Given the growing and significant interest in this area, we investigate the crowd-based research efforts in combating online misinformation.

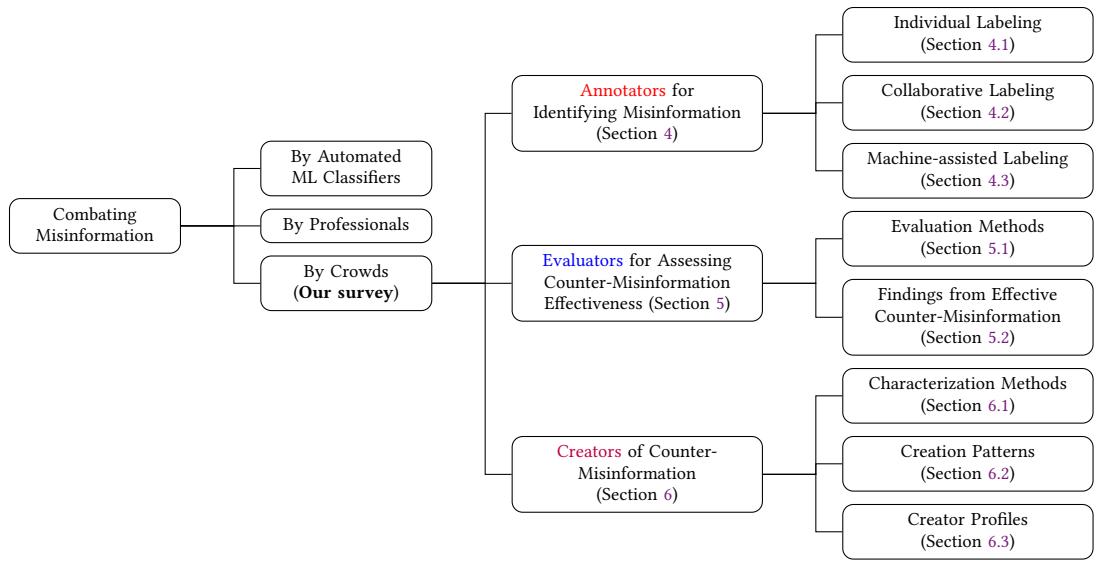
\begin{figure}
{\footnotesize
\begin{forest}
  for tree={
    grow'=0,
    draw,
    align=c,
    rounded corners,
    parent anchor=east,
    child anchor=west,
    l sep=30, 
    text centered, 
    text width=90,
    edge path={%
      \noexpand\path [\forestoption{edge}] (!u.parent anchor) -- ++(20pt,0) |- (.child anchor)\forestoption{edge label};
    }
  },
  [{Combating\\Misinformation},text width=60
      [{By Automated\\ML Classifiers}, l=5, text width=60]
      [{By Professionals},  l=5, text width=60]
      [{By Crowds\\(\textbf{Our survey})}, l=5, text width=60
      [{{\color{red}Annotators} for\\Identifying Misinformation\\(Section \ref{sec:help_detect_misinformation})}, 
      [{Individual Labeling\\(Section \ref{sec:4.1 individual labeling})}]
      [{Collaborative Labeling\\(Section \ref{sec:4.2 Collaborative Labeling})}]
      [{Machine-assisted Labeling\\(Section \ref{sec:4.3 Machine-assisted Labeling})}] 
      ]
      [{{\color{blue}Evaluators}  for Assessing\\Counter-Misinformation\\ Effectiveness (Section \ref{sec: crowds as evaluators})}, 
      [{Evaluation Methods\\(Section \ref{sec:evaluation methods})}]
      [{Findings from Effective\\Counter-Misinformation\\(Section \ref{sec:5.2 crowd evaluator findings})}]
      ]
      [{{\color{purple}Creators} of Counter-\\Misinformation\\ (Section \ref{sec:analyze_the_spread_of_existing_counter})}, 
      [{Characterization Methods\\(Section \ref{sec:Characterization Methods})}]
      [{Creation Patterns\\(Section \ref{sec:creation patterns})}]
      [{Creator Profiles\\(Section \ref{sec:creator profiles})}]
      ]
      ]
  ]
\end{forest}}
\caption{Two existing approaches to combat misinformation and our proposed taxonomy hierarchy related to crowd-based efforts.}
\label{fig:landscape}
\end{figure}

\subsection{{Our Work}}
In this survey, we aim to provide a comprehensive overview of the collaborative efforts made by crowds in combating misinformation.
To clarify our scope, we focus on online misinformation\footnote{The term ``misinformation'' in this survey is commonly represented as the terms ``fake news'', ``rumor'', ``false information'', ``false news'', and ``conspiracy theory.'' }, rather than all information or offline information. 
\new{
More specifically, our investigation centers on actively pursued efforts in \textit{combating misinformation} within this field.
}

\begin{figure}[t]
\centering
\includegraphics[width=0.85\linewidth]{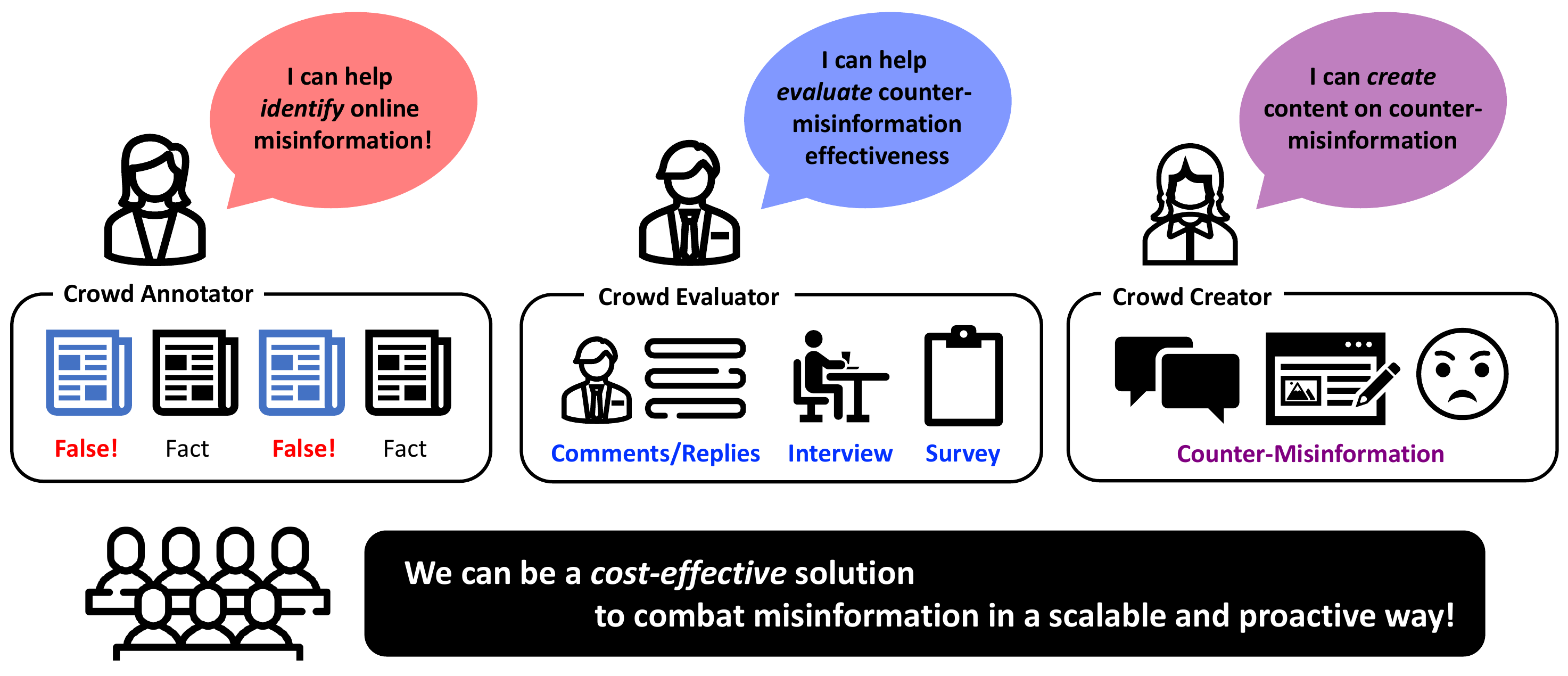} 
\caption{Illustration of the roles of crowds in combating misinformation.} \label{fig:roles}
\end{figure}

We first identify 88 relevant papers by following the guidelines of the preferred reporting items for systematic reviews and meta-analyses (PRISMA framework) \cite{moher2009preferred}  (Section~\ref{sec:collect_method}).
We then present a detailed overview of misinformation formats, topics, social media platforms, and crowd inputs (Section~\ref{sec:preli}). 
Upon holistic analysis of the papers, we propose a novel taxonomy of the crowd-based efforts, as depicted in Figure~\ref{fig:landscape}. 
To this end, we categorize crowd users based on their \textit{roles}, as shown in Figure~\ref{fig:roles}:

\begin{itemize} 
    \item \textbf{Crowds as Annotators:} Crowds help \textit{identify and label} online misinformation accurately at scale~\cite{Bhuiyan2020}, leveraging their extensive numbers compared to the limited professional fact-checker pool, and their widespread presence across social media platforms. 
    Additionally, crowds can \textit{amplify} the efforts of fact-checkers by sharing fact-checking articles out of a sense of social responsibility \cite{Pal2019}, driven by emotions such as anger or concern provoked by fact-checks \cite{Sun2021}, and a desire to warn others against misinformation \cite{Veeriah2021}.
    
    \item \textbf{Crowds as Evaluators:} Crowds help \textit{evaluate} the effectiveness and limitations of different counter-misinformation messages across various aspects.  They contribute by providing their first-hand experiences and tangible responses to misinformation and counter-misinformation.
    This valuable input facilitates the design of effective strategies to counter misinformation and offers a unique perspective not always accessible to professional fact-checkers~\cite{Borah2021}. 
    
    \item \textbf{Crowds as Creators:} Crowds \textit{create} posts on social media platforms to combat misinformation~\cite{Nadamoto2013, Chua2017}.
    They also respond to and comment on misinformation with accurate information \cite{guo2019future}.
    Analyzing such activities allows us to \textit{characterize} the creators and patterns of their counter-misinformation messages. 
\end{itemize}

We frame our examination of this topic around three research questions (RQs) that naturally align with the aforementioned roles of the crowds:

\begin{itemize}
    \item \textbf{(RQ1) Capabilities of crowds in identifying misinformation}: 
    How do crowds contribute to identifying or detecting misinformation, and how effective are they in this task?

    \item \textbf{(RQ2) Evaluation of counter-misinformation effectiveness by crowds}: 
    How can crowds be leveraged to evaluate counter-misinformation efficacy, and how effective are different types of counter-misinformation?

    \item \textbf{(RQ3) Characterization of counter-misinformation messages by crowds}: What are the characteristics of crowd-generated counter-misinformation messages and crowds who counter misinformation?

\end{itemize}

However, addressing the above RQs systematically poses several challenges. 
First, online misinformation spans a wide range of topics, including politics and natural disasters, and is disseminated across various platforms like Twitter and YouTube.
Meanwhile, crowds engage differently based on their roles, \new{including annotators, evaluators, and creators}.
Additionally, they counter misinformation across a range of content formats, including text and images. 
Lastly, researchers employ diverse approaches, such as in-lab experiments, interviews, and surveys, when analyzing crowd-based efforts in countering misinformation.

To navigate these challenges, our work to address each RQ can be summarized as follows:

\begin{itemize}

    \item \textbf{Crowds as {\color{red}annotators} for identifying misinformation} (for RQ1; Section~\ref{sec:help_detect_misinformation}): We investigate the capability of crowds to identify misinformation and compare it with that of professional fact-checkers. 
    We analyze individual and collective labeling scenarios, as well as the machine-assisted setting where humans and machines collaborate to enhance the annotation performance.

    \item \textbf{Crowds as {\color{blue}evaluators} for assessing the effectiveness of counter-misinformation} (for RQ2; Section \ref{sec: crowds as evaluators}): 
    We examine direct and indirect evaluation methods, including expressed sentiment and stance, in-lab experiments, interviews, and surveys, to \new{quantify} the efficacy of a given counter-misinformation message. Then, we investigate the distinct advantages offered by each counter-misinformation approach.

    \item \textbf{Crowds as counter-misinformation {\color{purple}creators}} (for RQ3; Section~\ref{sec:analyze_the_spread_of_existing_counter}): 
    We examine two aspects of counter-misinformation characteristics: 
    i) the characteristics of counter-misinformation messages generated by the crowds on social media platforms; 
    ii) typical attributes of crowds who counter misinformation. 
\end{itemize}

Through these investigations, we provide valuable insights into the effectiveness and limitations of existing crowd-based efforts in combating misinformation.

\begin{table}[!t]
\centering
\small
\caption{Comparison with existing surveys. In this table, `\cmark' and '\dashcheckmark' denote that the relevant survey fully and partially covered the corresponding topic, respectively.}
\label{tab:survey_comparison}

\renewcommand{\arraystretch}{1.1}
\begin{tabular}{cccc}  % Changed cccccccc to cccc
\toprule
& \multicolumn{3}{c}{\textbf{Crowds' Roles}} \\  
\cmidrule{2-4}  
& \begin{tabular}[c]{@{}c@{}}\textbf{{\color{red}Annotators}} for identi-\\{fying misinformation}\end{tabular} & \begin{tabular}[c]{@{}c@{}}\textbf{{\color{blue}Evaluators}} for assessing\\{counter-misinformation}\end{tabular} & \begin{tabular}[c]{@{}c@{}}\textbf{{\color{purple}Creators}} of\\{counter-misinformation}\end{tabular} \\ 
\midrule
\cite{wang2019systematic, suarez2021prevalence}    &    &    &     \\
\cite{Sharma2019, guo2019future, Shuhud2017}    &  \dashcheckmark    &    &     \\
\cite{chan2017debunking, hartwig2023landscape}    &  &  \cite{chan2017debunking}: \cmark, \cite{hartwig2023landscape}: \dashcheckmark    &     \\
\textbf{Our work}    &  \cmark & \cmark & \cmark    \\
\bottomrule
\end{tabular}
% }
\end{table}

\subsection{{Related Survey}}

While most surveys on misinformation  focus on automated machine learning solutions~\cite{islam2020deep}, \new{it is} worth noting that a few surveys \cite{Sharma2019, Shuhud2017, guo2019future,chan2017debunking,hartwig2023landscape} \new{cover crowd-based efforts}.
However, these previous surveys have certain limitations, and we aim to provide a more comprehensive perspective. Here's a comparison between our survey and previous ones, as shown in Table \ref{tab:survey_comparison}.

First, some surveys \cite{Sharma2019, Shuhud2017, guo2019future} have examined crowds' ability to \textit{identify misinformation} (\ie, {\color{red}annotators}). 
They \new{infer} misinformation-related signals from indirect crowd behaviors, such as replies and comments, and \new{incorporate} those signals into ML solutions as labels. 
However, as mentioned in Section~\ref{sec:motivation}, our survey recognizes and investigates the potential for crowds to serve as a direct and potent means for accurately identifying and labeling misinformation at a large scale. This distinction sets our survey apart from the prior research.
Second, \citet{chan2017debunking} summarize the effectiveness of various types of counter-misinformation \textit{evaluated by crowds} (\ie, {\color{blue}evaluators}). 
However, this survey has limitations as it only covers 8 papers, is restricted to literature published before 2018, and does not \new{explore} the evaluation methods employed by crowds. 
\citet{hartwig2023landscape} analyze the user-centered misinformation interventions where crowds implicitly or explicitly evaluate certain intervention techniques. Nevertheless, some interventions are not related to counter-misinformation contents, \eg, the removal of misinformation.
In contrast, our survey addresses these limitations through a rigorous paper search process for recent papers and summarizes the metrics and methods used to evaluate the efficacy of counter-misinformation.
Third, none of the existing surveys cover the \textit{characterization of counter-misinformation} from the perspective of the crowds (\ie, {\color{purple}creators}).
This is a notable gap as understanding the characteristics of crowd-generated counter-misinformation can offer valuable insights for devising effective strategies to combat misinformation. 
Our survey fills this gap by \new{comprehensively analyzing}  
the patterns of existing counter-misinformation messages generated by crowds, \new{and} identifying and profiling typical attributes of these crowd creators.

\subsection{Contributions}
In sum, the main contributions of this survey are:
\begin{itemize}

    \item \textbf{Comprehensive Survey:} We systematically identify the relevant papers on the crowd-based efforts in combating misinformation and then review them \new{regarding} detection of misinformation, evaluation of counter-misinformation effectiveness, and characterization of counter-misinformation creation.  
    To the best of our knowledge, this is the first review of the literature that encompasses crowds' contributions to these three crucial aspects.

    \item \textbf{Key Statistics:} We summarize important data statistics regarding misinformation and crowds found in the literature.
    This \new{contains} the common formats and topics of misinformation, the social media platforms where crowds engage, and the inputs made by the crowds.
    
    \item \textbf{Novel Taxonomy:} We provide a novel taxonomy of approaches that comprehensively covers the diverse functions of crowds.
    This is designed to help researchers understand the current research trends in this area.

     \item \textbf{Thorough Analysis:}  We conduct a comprehensive analysis of each crowd role, including individual, collaborative, and machine-assisted labeling for annotators, survey, interview, and in-lab experiment-based counter-misinformation effectiveness assessment for evaluators, and the characterization of creation patterns and creator profiles for creators.
     
    \item \textbf{Future Directions:} We discuss the limitations of existing crowd-based approaches to combat misinformation and suggest several promising research directions for the future.
    
\end{itemize}

\section{Method of Identifying Relevant Papers}\label{sec:collect_method}

Following PRISMA guidelines \cite{moher2009preferred}, we conducted a comprehensive search  for relevant papers on \textit{www.scopus.com}, a reputable scientific database. 
The following query was executed on September 22, 2022, and yielded 3,956 papers. 
\begin{figure}[H]\fontsize{8pt}{8pt}
\centering
\begin{BVerbatim}[commandchars=\\\{\}]
TITLE-ABS-KEY ( (Category \#1) AND  ( Category \#2 ) AND  ( Category \#3 ) AND  ( Category \#4 ) )  
AND  PUBYEAR  >  1999  AND  ( LIMIT-TO ( LANGUAGE,  "English" ) ) 
\end{BVerbatim}
\end{figure}
We specifically targeted research articles in English published after 1999.
The search utilized information from titles, abstracts, and keywords of these articles, referred to as ``\texttt{TITLE-ABS-KEY}''. The search combined four categories of information using logical ``\texttt{AND}''.
These categories and their definitions are provided in Table \ref{tab:keyword_set}. 
Keywords within each category were combined using the ``\texttt{OR}'' operator to ensure that all related concepts were included, and the wildcard ``\texttt{*}'' was used to account for multiple spelling variations.
Specifically, Category 1 deals with terms related to crowds, particularly non-experts, engaged in countering misinformation. 
Category 2 is related to online platforms where crowd reactions to misinformation are observed.
Category 3 covers misinformation-related terms and synonyms.
Lastly,  Category 4 indicates actions that hinder the spread of misinformation.

\begin{table}
    \centering
    \caption{The collection of keywords used to search relevant papers.}
    \label{tab:keyword_set}
    \begin{tabular}{cp{12cm}}
    \toprule
        \textbf{Categories} & \textbf{Keywords} \\ \midrule
        \multirow{2}{*}{Category 1}   & \texttt{crowd  OR  citizen  OR  community  OR  crowdsourcing  OR  group  OR  user  OR  people  OR  society  OR  human  OR  individual}  \\ \midrule
        \multirow{2}{*}{Category 2} &  \texttt{web  OR  "social media"  OR  "social network"  OR  internet  OR  online  OR  twitter  OR  facebook  OR  instagram  OR  whatsapp  OR  weibo  OR  wechat  OR  reddit  OR  tumblr}  \\ \midrule
        \multirow{2}{*}{Category 3}  & \texttt{misinformation  OR  fake  OR  misleading  OR  disinformation  OR  conspiracy  OR  rumors  OR  "false information"  OR  hoax  OR  *infodemic} \\ \midrule
        \multirow{3}{*}{Category 4} & \texttt{counter*  OR  fight*  OR  respond*  OR  argu*  OR  negate*  OR  "fact check*"  OR  reply*  OR  dispute*  OR  respond*  OR  refute*  OR  debunk*  OR  flag*  OR  judg*  OR  combat*  OR  censor*  OR  correct*  OR  block*} \\ \bottomrule
    \end{tabular}
\end{table}

To ensure the inclusion of only relevant papers in our survey, we rigorously followed PRISMA guidelines and established the following inclusion criteria:
First, the paper must explicitly mention \textit{crowds' active community engagement}. This involvement typically encompasses crowd inputs such as annotations, replies to, or comments on online misinformation; or responses related to counter-misinformation collected through surveys, in-lab experiments, and interviews.
Second, the purpose of the aforementioned \old{efforts} \new{crowd engagements} or the associated research papers should be \textit{combating misinformation}.
This implies that the paper should employ these \old{efforts} \new{crowd engagements} to mitigate the negative impacts of misinformation and promote the positive effects of counter-misinformation \new{for combating misinformation}. 
These actions may include detecting misinformation, evaluating counter-misinformation effectiveness, and characterizing counter-misinformation.

\begin{figure}[!htbp]
    \centering
    \includegraphics[width=0.43\textwidth]{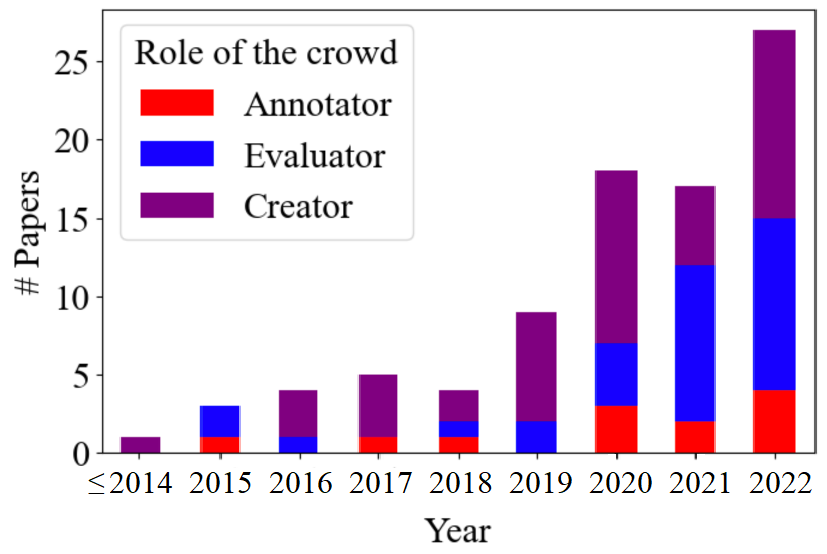} 
    \caption{Annual distribution of annotated relevant papers in our survey categorized by the role of the crowd. }
    \label{fig:example}
\end{figure}

These criteria were established through extensive discussions between two authors to ensure the selection of papers within the intended topic. 
Each author initially assessed a batch of 200 papers \new{among the total of 3,956 papers} together \new{to establish the annotation criteria and check the inter-rater agreement score}.
Papers were categorized as ``Yes'' if they explicitly met both criteria,
``No''  if they didn't focus on the research topic or were proposal articles, and ``Maybe'' if there was some degree of confidence in their relevance.
Subsequently, an inter-rater reliability analysis was conducted, yielding a Krippendorff's alpha score \cite{krippendorff2011computing} of 0.571, indicating a moderate level of agreement~\cite{mchugh2012interrater}. 
This seemingly low but acceptable value accounts for some overlap between ``Maybe'' and ``Yes''  labels \new{for the same papers}, which was resolved during the collaborative review. 
\new{Particularly, in the collaborative review, two authors discuss and finalize the label for the confusing ``Maybe'' papers after reading the full paper}. 
\new{We eventually had 19 ``Yes'' papers and 181 ``No'' papers for the initial batch of 200 papers.} 
\new{Next, we repeat the process for the remaining 3,756 papers where each author annotated 1,878 papers separated and finally had 40 ``Yes'', 121 ``Maybe'', and 3,595 ``No'' papers. Likewise, two authors had the collaborative review for these 121 ``Maybe'' papers and identified 29 ``Yes'' papers. 
} 
Ultimately, this comprehensive process identified 88 papers for our survey \new{by combining the results of the first, second, and third passes --  the first pass identified 19 ``Yes'' papers, the second identified 40 papers, and the third identified 29 papers.} 
Figure \ref{fig:example} displays the annual distribution of our selected papers categorized by the role of the crowd.

\section{Data Statistics of Papers}\label{sec:preli}

We \old{provide a summary of} summarize the selected papers in the survey, examining relevant statistics regarding the formats of misinformation, covered topics, utilized social media platforms, and crowd inputs. This overview offers scholars a preliminary understanding of the research field.

\subsection{Misinformation} 

\begin{table}[t]
    \centering
    \caption{Overview of misinformation statistics in surveyed papers.}
    \label{tab:counter_misinformation_stat}
    \renewcommand{\arraystretch}{1.1}
    \begin{tabular}{llll} 
        \toprule
         & \multicolumn{2}{c}{\textbf{Categories}} & \textbf{References} \\
         \midrule
        \multirow{6}{*}{{\rotatebox{90}{\textbf{(a) Topics}}}} &\multicolumn{2}{c}{Politics}  &    \cite{Chua2017, Masullo2021, Cohen2020, Chua2017study, Goh2017, Shabani2018, Soprano2021, li2020social, Dang2016, Martel2021} \\
        &\multicolumn{2}{c}{Natural Disasters} & \cite{Nadamoto2013, Flores2021, Bhuiyan2020, Hunt2020, Wang2018, Weber2020, li2020social}\\
        &\multicolumn{2}{c}{Health Issues} & \cite{Gunaratne2019, Stojanov2015, Featherstone2020, Kim2021, Vraga2021, Meer2020, Masullo2021, Allen2021, Pal2019, Pal2018, Memon2020, Ahmed2020, Micallef2020, Igbinovia2021, Borah2021, Veeriah2021, Mohammad2021, Sun2021, Bode2018, xue2022covid, kim2022use} \\
        &\multicolumn{2}{c}{Crisis Events} & \cite{Zeng2019, Wang2018, McCreadie2015, Jung2020, Lee2021, Babcock2019, Arif2017, Zhao2016, Zubiaga2016} \\
        &\multicolumn{2}{c}{Civic Topics} & \cite{Pal2018, Pal2017, Tanaka2019, Mancosu2021, Chua2017, Chua2017study, Allen2021, li2020social, Pennycook2019, Babcock2019, Babcock2019JDIQ, Vafeiadis2019, Franklin2020} \\
        &\multicolumn{2}{c}{General Topics} & \cite{Vo2019, Kirchner2020, Mitra2015, Ramachandran2020, Pal2019, giachanou2022impact} \\
        \midrule
        
    \multirow{11}{*}{\rotatebox{90}{\textbf{(b) Platforms}}} &\multirow{8}{*}{Social Media} & Twitter & \cite{Nadamoto2013, Lee2021, Chua2017study, Gunaratne2019, Zubiaga2016, Micallef2020, Babcock2019, Ahmed2020, Flores2021, Hunt2020, Jung2020, Goh2017, Babcock2019JDIQ, Arif2017, Memon2020, Wang2018, Weber2020, Pal2018, Pal2017, Shabani2021, giachanou2022impact, buchanan2022reading} \\
    && Facebook & \cite{Mohammad2021, Allen2021, Flores2021, Shabani2021, xue2022covid}\\
    && YouTube & \cite{kim2022use, buchanan2022reading} \\
    && Reddit & \cite{Dang2016, Achimescu2020} \\
    && Sina Weibo & \cite{Zeng2019, li2020social} \\
    && Zhihu & \cite{chen2022science}\\
    && Whatsapp & \cite{kligler2022collective} \\ 
    \cmidrule(lr){2-4}
 
    &\multirow{2}{*}{Crowdsourcing} & AMT &  \cite{Featherstone2020, Pennycook2019, Soprano2021, Martel2021, Sun2021} \\
    && {Others} & \cite{Orosz2016, Stojanov2015, Mancosu2021, Meer2020, Sun2020, Igbinovia2021, Bhuiyan2020, Borah2021, Veeriah2021, Pundir2021, Zhao2016, Masullo2021, Cohen2020, Vraga2021, Mitra2015, Tanaka2019, Kim2021,  Bode2018, McCreadie2015, Vafeiadis2019, Pal2019, Kirchner2020, Shabani2018} \\ 
    \cmidrule(lr){2-4}
    
    & \multicolumn{2}{c}{Other Platforms} & \cite{Ramachandran2020, Pal2019} \\ \bottomrule
    \end{tabular}
\end{table}

\subsubsection{Formats and Topics}

Countered misinformation mainly comprises textual content \new{e.g., posts on social media platforms} \old{encompassing textual posts on social media platforms}. 
Research has explored a diverse range of misinformation topics, as summarized in Table~\ref{tab:counter_misinformation_stat}-(a).
These topics consist of politics (e.g., elections \cite{Cohen2020} and immigration \cite{Masullo2021}); natural disasters (e.g., earthquakes \cite{Nadamoto2013} and climate change \cite{Bhuiyan2020}); 
health issues (e.g., COVID-19 pandemic \cite{Micallef2020, xue2022covid, kim2022use, buchanan2022reading}, vaccines \cite{Stojanov2015, Gunaratne2019, xue2022covid}, and genetically modified organisms~\cite{chen2022science}); crisis events (e.g.,  mass shooting \cite{Lee2021}); 
and other civic subjects, including rumors about brands KFC \cite{Pal2018, Pal2017}, celebrities \cite{Chua2017, Chua2017study}, and movies \cite{Babcock2019, Babcock2019JDIQ}.
The research also addresses generic misinformation topics obtained from online fact-checking sources like \textit{Snopes.com} and \textit{PolitiFact.com} \cite{Vo2019, Kirchner2020}.

\subsubsection{Platforms}
The crowds actively combat misinformation across various online platforms \old{where they} primarily \old{work} as counter-misinformation creators, including social media platforms like Twitter and Facebook. 
Additionally, crowd-sourcing platforms like Amazon Mechanical Turk (AMT) are leveraged to collect annotation of misinformation and evaluation of counter-misinformation \new{effectiveness} \old{from crowds}. 
A summary of these platforms can be found in Table~\ref{tab:counter_misinformation_stat}-(b).

\begin{table}[t]
    \centering
    \caption{Overview of crowd inputs statistics in surveyed papers.} 
    \label{tab:counter_misinformation_stat_crowd}
    \renewcommand{\arraystretch}{1.1}
    \begin{tabular}{clll} 
        \toprule
         \multicolumn{3}{c}{\textbf{Categories}} & \textbf{References} \\
         \midrule
        % \multirow{3}{*}{\rotatebox{90}{\textbf{(a) Formats}}} 

       \multirow{3}{*}{\rotatebox{90}{\begin{tabular}{@{}c@{}}\textbf{(a)}\\ \textbf{Format}\end{tabular}}}
        && \multicolumn{1}{l}{Text} &   \cite{Zeng2019, Ahmed2020, Flores2021, Jung2020, Goh2017, Arif2017, Memon2020, Wang2018, li2020social, Pal2017, xue2022covid, giachanou2022impact, Micallef2020, chen2022science, kligler2022collective} \\ 
        && \multicolumn{1}{l}{Image} &  \cite{Dang2016, Pal2017, Chua2017study, buchanan2022reading} \\ 
        && \multicolumn{1}{l}{Video} &   \cite{kim2022use} \\  
        % \\[1pt]
        \midrule

    \multirow{17}{*}{\rotatebox{90}{\textbf{(b) Features}}} &\multirow{4}{*}{Explicit} & Flagging  Misinformation &         \cite{mujumdar2021hawkeye, allen2022birds} \\
    && Credibility of News & \cite{Bhuiyan2020, Mitra2015, Allen2021, McCreadie2015, Vraga2021, Orosz2016, Mancosu2021, Meer2020, Kim2021, Kirchner2020, Bode2018, Martel2021, Vafeiadis2019} \\
    && Debunking Websites & \cite{Micallef2020, Veeriah2021, Allen2021, Pennycook2019, McCreadie2015} \\
    && Countermeasures & \cite{Ramachandran2020, Shabani2018, Kirchner2020, Featherstone2020, Masullo2021, Martel2021, Vafeiadis2019, Bode2018, Kim2021, Meer2020, Cohen2020, Stojanov2015, Sun2021, Pal2018, Pal2019} \\
    \cmidrule(lr){2-4}
    % \addlinespace[8pt]

    & \multirow{9}{*}{Implicit} &  Textual Embedding & \cite{Micallef2020} \\
    & & Psycholinguistic Features & \cite{Micallef2020, Shabani2018, Memon2020, Pal2018, Chua2017study, Gunaratne2019, xue2022covid, giachanou2022impact} \\
    & & Topic & \cite{Shabani2018, Dang2016} \\
    & & Sentiment & \cite{Pal2018, Micallef2020, Shabani2018, li2020social, Dang2016} \\
    && Emotion & \cite{kim2022use} \new{\cite{Sun2020, Pundir2021, Meer2020}} \\
     && \new{Demographics} & \new{\cite{Pal2018, Pal2019,Igbinovia2021, Zhao2016}} \\
     && \new{Media Literacy} & \new{\cite{Vraga2021, Veeriah2021, Pundir2021}} \\
     && \new{Conspiracy Mentality} & \new{\cite{Mancosu2021, Orosz2016, Bode2018}} \\ 
    && Hashtag & \cite{Ahmed2020, Babcock2019JDIQ, Weber2020} \\
    && URL  & \cite{Chua2017study, Ahmed2020, Hunt2020, Jung2020, Weber2020, Micallef2020} \\
    && Number of Likes/Shares & \cite{Hunt2020,Pal2017} \\
    && Group Identity Language & \cite{chen2022science} \\
    \bottomrule
    \end{tabular}
\end{table}

\subsection{Crowd Inputs}

\subsubsection{Content Formats}

Crowds counter misinformation through diverse content formats, as outlined in Table~\ref{tab:counter_misinformation_stat_crowd}-(a). 
The primary involves utilizing various textual formats such as posting counter-misinformation content, commenting on news articles, replying to social media posts, and retweeting or sharing corrective information~\cite{Zeng2019, Ahmed2020, Flores2021, Jung2020, Goh2017, Arif2017, Memon2020, Wang2018, li2020social, Pal2017}. 
Additionally, images often supplement textual content to enhance the effectiveness of countering misinformation \cite{Pal2017, Chua2017study}. 
Lastly, video content serves as an effective tool to debunk misinformation on platforms like YouTube and has demonstrated its potential to educate the general public \cite{kim2022use}.

\subsubsection{Features Extracted from Crowd Inputs}
Crowds provide a diverse range of content in response to misinformation, offering researchers valuable features that can be utilized for misinformation detection and counter-misinformation characterization. 
In this survey, we categorize them into explicit and implicit features, as shown in Table~\ref{tab:counter_misinformation_stat_crowd}-(b). 
Explicit features involve a direct examination of raw inputs.
This includes activities such as rating and flagging misinformation~\cite{mujumdar2021hawkeye, allen2022birds}, assessing credibility scores of news articles~\cite{Bhuiyan2020, Mitra2015, Allen2021, McCreadie2015, Vraga2021, Orosz2016, Mancosu2021, Meer2020, Kim2021, Kirchner2020, Bode2018, Martel2021, Vafeiadis2019}, identifying valuable debunking websites~\cite{Micallef2020, Veeriah2021, Allen2021, Pennycook2019, McCreadie2015}, suggesting countermeasures~\cite{Ramachandran2020, Shabani2018, Kirchner2020, Featherstone2020, Masullo2021, Martel2021, Vafeiadis2019, Bode2018, Kim2021, Meer2020, Cohen2020, Stojanov2015, Sun2021, Pal2018, Pal2019}.
Implicit features, on the other hand, are derived through \new{applying computation methods to raw inputs or giving questionnaires to crowds}. 
These \new{computational} methods generate new feature vectors, such as textual embeddings~\cite{Micallef2020}, psycholinguistic features \cite{Micallef2020, Shabani2018, Memon2020, Pal2018, Chua2017study, Gunaratne2019, xue2022covid, giachanou2022impact}, sentiment~\cite{Pal2018, Shabani2018, li2020social, Dang2016}, and other computational metrics 
~\cite{li2020social, Babcock2019JDIQ, Chua2017study, Ahmed2020, Hunt2020, Jung2020, Weber2020, Micallef2020,Pal2017,chen2022science} for (counter-)misinformation analysis. 
\new{ 
Additionally, the user-related implicit features from answered questionnaires by crowds contain demographic information ~\cite{Pal2018, Pal2019} (e.g., age and education~\cite{vijaykumar2022dynamics}), emotion~\cite{Sun2020, Pundir2021, Meer2020} (e.g., fear of missing out~\cite{Pundir2021}), media literary indicating the ability to critically analyze and evaluate various news information~\cite{Vraga2021, Veeriah2021, Pundir2021}, conspiracy mentality~\cite{Mancosu2021, Orosz2016, Bode2018},  the mindset to generally believe in conspiracy theories, and group identity language~\cite{chen2022science} where people who share social factors and belong to the same social group are more likely to use the similar language terms.
}

\section{Crowds as Annotators for Identifying Misinformation}\label{sec:help_detect_misinformation}

Following our meta-level summary of relevant papers, we now dive deeper into the functional categorization of the roles of crowds. Our taxonomy, represented in Figure~\ref{fig:roles}, categorizes the contributions of crowds along three axes: annotators, evaluators, and creators. In this section, we focus on the ``annotator''  category, where crowds serve as annotators for identifying misinformation.

\begin{figure}[t]
\centering
\includegraphics[width=0.95\linewidth]{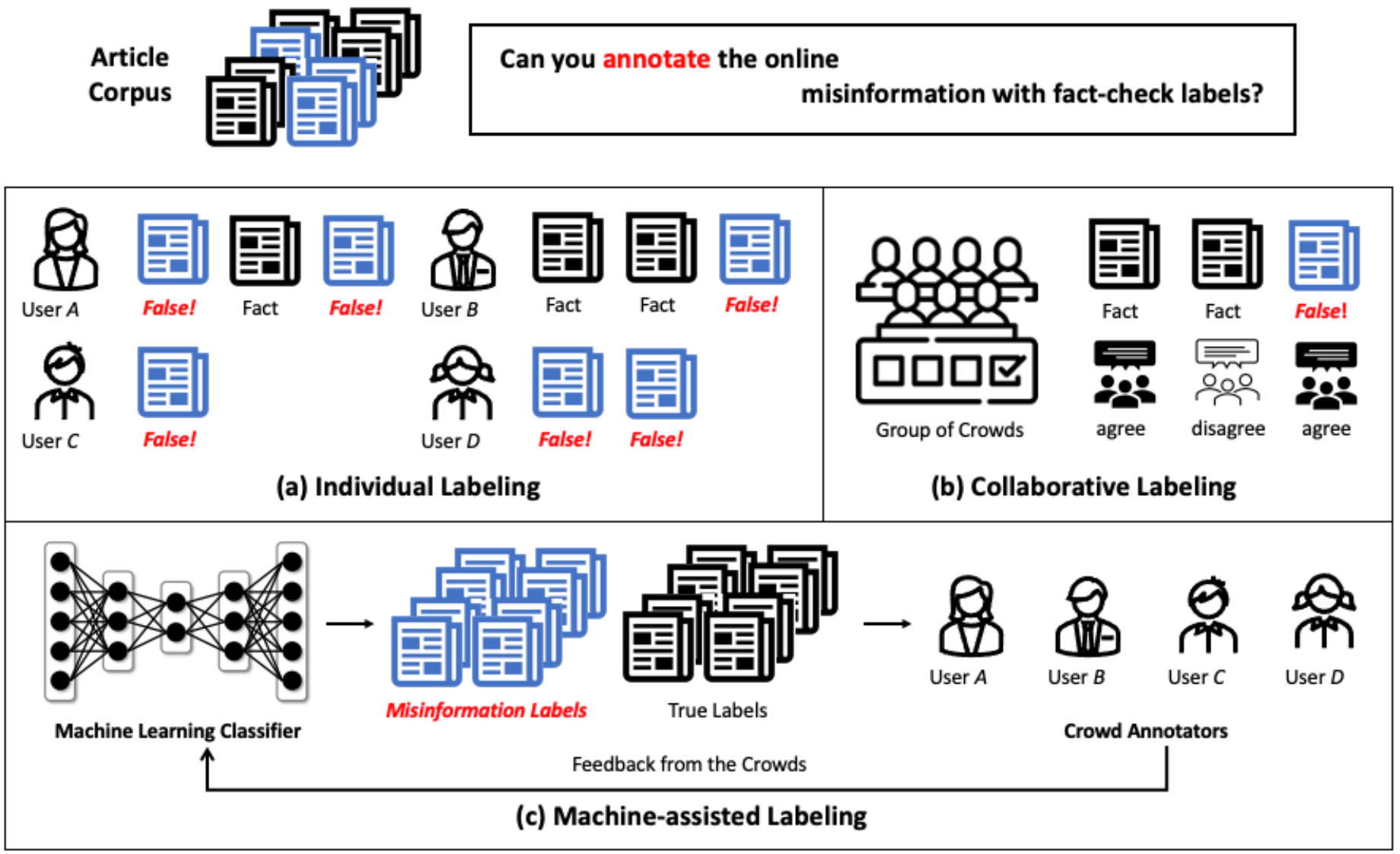} 
\caption{Illustration of crowds' role as annotators for identifying misinformation.}
\label{fig:annotators}
\end{figure}

To identify misinformation, many ML methods have emerged for automated detection \cite{Sharma2019}.  
However, these methods heavily rely on the expertise of fact-checkers to obtain ground-truth annotations (\eg, fact-check labels), whose population and bandwidth are unfortunately limited. 
In contrast, emerging approaches \old{involve harnessing} \new{harness} the fact-checking competence of the crowd, especially laypeople, for efficient and effective misinformation annotation \cite{Ramachandran2020,  Mitra2015}. 
In this context, crowds can directly label potential misinformation articles. 
Notably, recent research suggests that crowd-based annotation performance rivals that of professionals \cite{Bhuiyan2020, roitero2021can}, enabling the use of these labels in downstream tasks such as misinformation classification \cite{Shabani2018} and (counter-)misinformation analysis \cite{Goh2017}. 
Therefore, we review the relevant studies where crowds annotate misinformation.  These studies fall into three key categories: individual labeling of misinformation, collaborative labeling by a group of \old{annotators} \new{users}, and machine-assisted labeling with crowd inputs. Figure~\ref{fig:annotators} provides a visual representation, and Table \ref{tab:Misinformation Detection by Crowd} offers a \old{comprehensive} list of reference papers. 
We then provide detailed descriptions of each category in the subsequent subsections.

\begin{table}[t]
    \centering
    \caption{Taxonomy of crowd annotators for identifying misinformation.}
    \label{tab:Misinformation Detection by Crowd}

    \begin{tabular}{ll} \toprule
     \textbf{Categories} & \textbf{References} \\ \midrule
    
    Individual Labeling  &    \cite{roitero2021can,Ramachandran2020, Mitra2015, xu2022does,Bhuiyan2020} \\ 
    
    Collaborative Labeling  &  \cite{mujumdar2021hawkeye, allen2022birds} \\ 
    % \cite{Shabani2018, Mitra2015}
    
    Machine-assisted Labeling  &   \cite{Shabani2018, Mitra2015, Ramachandran2020, farooq2022crowd} \\ \bottomrule
    \end{tabular}

\end{table}

\subsection{Individual Labeling}
\label{sec:4.1 individual labeling}

Individual labeling of misinformation requires independent judgments about the credibility of the given misinformation by each crowd member. 
To \old{accomplish this labeling} \new{achieve it}, platforms such as Amazon Mechanical Turk \cite{roitero2021can} and Upwork \cite{Bhuiyan2020} \new{are used to} facilitate direct user labeling \cite{Ramachandran2020, Mitra2015}.  
\old{On these platforms,} Labels obtained from crowds are typically verified through majority voting.
During this process, the credibility levels of the crowds increase as they accurately identify misinformation.

Furthermore, several factors \old{have been explored that} influence the quality of individual labeling \cite{xu2022does,Bhuiyan2020,roitero2021can}.
One significant factor is {evidence from other peers}~\cite{xu2022does}, which can either aid or mislead \old{crowds in their judgments} \new{crowds' judgments}. 
Crowds who effectively use provided evidence tend to make more accurate annotations. 
Moreover, the {demographic and political composition} can \old{also} influence crowds' credibility ratings and annotations. 
For example, \citet{Bhuiyan2020} found that Democrats, males, those between the ages of 26 and 30, and those with higher levels of education are more likely to agree with experts on climate science. 

Additionally, {characteristics of the annotation task} itself, such as 
the genre of the article and partisanship of the publication 
, 
also have an impact. 
\citet{Bhuiyan2020} found that crowds demonstrate a higher correlation with experts in opinion articles and left-leaning publications.
Lastly, the {length of the annotation period} can also affect the quality of the judgments \old{made by crowds}. 
Notably, \citet{roitero2021can} found that annotations collected at different time spans for the same document may yield different results.
\new{Instead,} annotations collected \old{in close proximity to } \new{close to} each other tend to produce similar outcomes.

\subsection{Collaborative Labeling}
\label{sec:4.2 Collaborative Labeling}

In addition to individual labeling, collaborative labeling indicates that multiple individuals \old{are contributing} \new{can contribute} to annotations collaboratively, which enhances quality through shared insights and community engagement \cite{mujumdar2021hawkeye, allen2022birds}. 
Twitter's Birdwatch/Community Notes\footnote{\url{https://help.twitter.com/en/using-twitter/community-notes}}, introduced in 2021, is a prominent example.
In this initiative, crowds review the annotations of others and then label tweets that may contain misinformation by providing supporting material \old{ and annotations of others}.
It is worth noting that, unlike majority-based aggregation commonly used in individual labeling scenarios, collaborative labeling involves group-level interactions that occur \textit{before} final label consensus. This aspect makes collaborative labeling a more crowd-enabled process~\old{compared to individual labeling}.
\old{Subsequently, research efforts have emerged to} Subsequent research efforts investigate and address challenges within the Birdwatch collaborative labeling ecosystem. 
For instance, crowds have different levels of credibility in their annotations, which can result in biased or unfair labeling. 
To tackle this issue, \citet{mujumdar2021hawkeye} proposed HawkEye, a robust reputation system for fair user ranking and misinformation labeling.
Additionally, \citet{allen2022birds} investigated the impact of crowds' partisanship~\old{within the Birdwatch annotation process}. They found that crowds tend to offer negative annotations for tweets from counter-partisans and consider their annotation less helpful.

\subsection{Machine-assisted Labeling}
\label{sec:4.3 Machine-assisted Labeling}

Machine-assisted labeling methods combine computational power and crowd inputs to efficiently detect misinformation \cite{Shabani2018, Mitra2015}. 
In the human-in-the-loop pipeline proposed by \citet{Shabani2018}, crowds initially verify misinformation labels assigned by ML classifiers, and this feedback is then used to refine the classifiers~\cite{Shabani2018}. 
Additionally, \citet{farooq2022crowd} proposed a blockchain-based framework that leverages the inherent immutability and incentive features of blockchains to record crowd annotations, ensuring accountability while also rewarding accurate contributions and penalizing malicious annotations. 
Similarly, \citet{Ramachandran2020} incorporated the blockchain framework to collect the given genuineness score of news articles from crowds to improve the downstream fake news detection algorithms. 
Lastly, topic modeling techniques can be used to cluster tweets on similar topics, and crowds can assess the credibility of these tweets to ensure a coherent and consistent annotation process \cite{Mitra2015}.

\section{Crowds as Evaluators for Assessing Counter-Misinformation Effectiveness} \label{sec: crowds as evaluators}

\begin{figure}[t]
\centering
\includegraphics[width=0.96\linewidth]{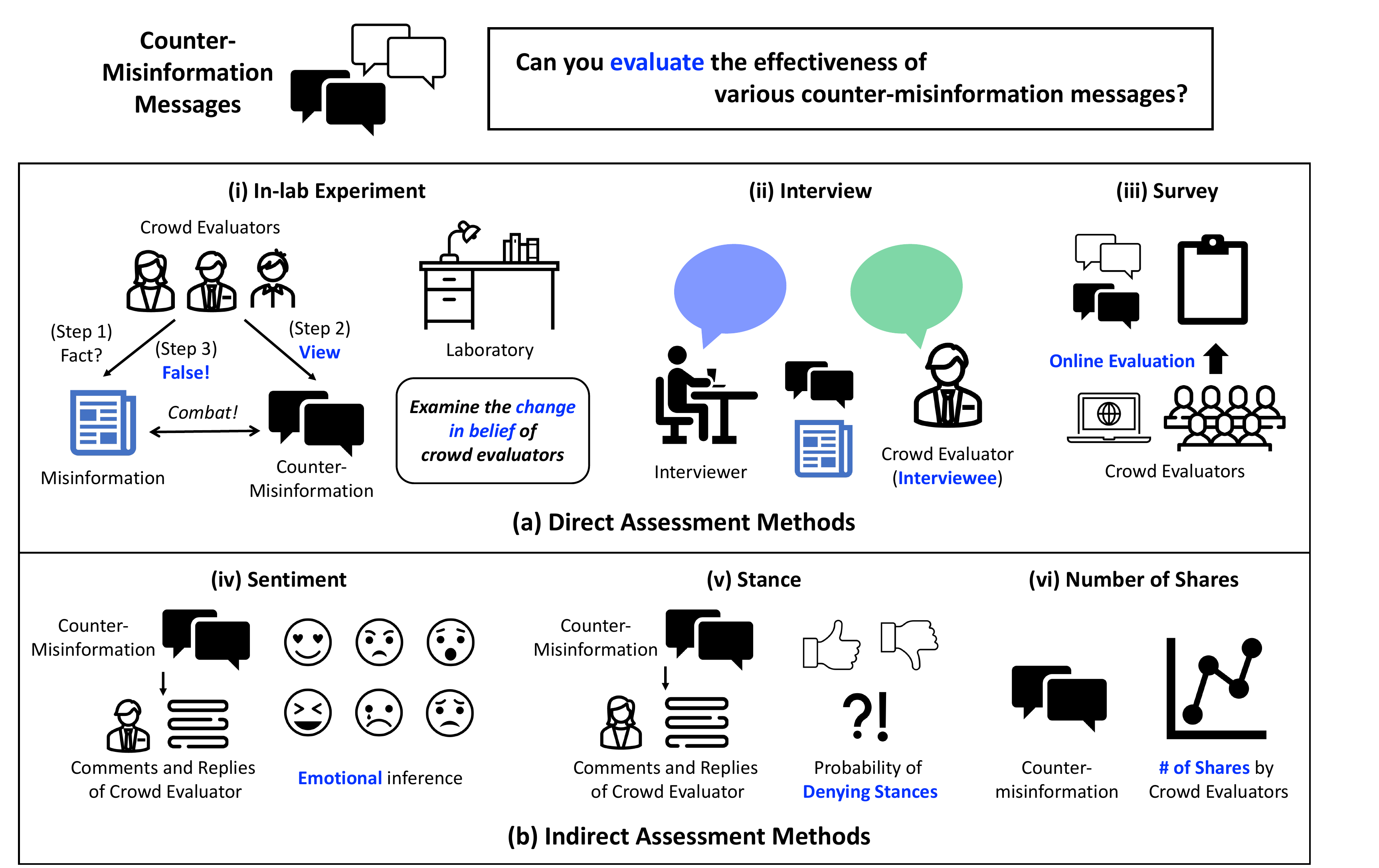} 
\caption{Illustration of crowds' role as evaluators to assess counter-misinformation effectiveness.}
\label{fig:evaluators}
\end{figure}

Once misinformation is identified by crowds, ML classifiers, or professionals, different counter-misinformation messages are created on social media platforms \new{for debunking}. 
In this case, crowds can help evaluate the effectiveness and limitations of various counter-misinformation messages.
To introduce evaluation methods utilizing crowds as evaluators in the literature, we \old{categorize} \new{group} them into two categories, \ie, direct and indirect assessments \old{by crowds} (see Figure~\ref{fig:evaluators}). 
Each category consists of three sub-categories: direct assessment--in-lab experiment, interview, and survey; indirect assessment--sentiment, stance, and \new{the} number of shares.
Next, we describe the details of each category and \old{then} discuss findings regarding effective counter-misinformation across different factors, such as \old{media formats, countering order, word placement, content traits, communication styles, and audience-related factors} \new{media formats, content traits, communication styles, audience factors, countering order, and word placement}, in the subsequent subsections.
The taxonomy of evaluation methods and factors can be found in Table~\ref{tab:Evaluation Metrics and Methods}-(a) and -(b), respectively.

\subsection{Crowd-based Evaluation Methods}\label{sec:evaluation methods}

As mentioned above, two \old{evaluation methods} \new{paradigms} can be employed to evaluate the efficacy of counter-misinformation using \old{crowd input} \new{crowds}. 
The first method, referred to as \textit{direct assessment}, involves \old{the use of direct quantitative questions in which crowds rate the believability of misinformation after encountering counter-misinformation messages} \new{using quantitative questions where crowds rate the believability of misinformation after viewing counter-misinformation messages}.  
This approach directly assesses the preferences and perceptions of the crowd, with lower believability indicating more effective countermeasures. 
In contrast, the second method, termed \textit{indirect assessment}, involves \old{the use of } using \old{indirect indicators or} proxy metric. 
\old{To this end, we can analyze attitudes expressed in the comments and fact-checking shares of the crowds to indirectly measure the effectiveness of countering misinformation.} 
\new{Particularly, we can analyze the sentiment, stance, or the number of shares of counter-misinformation to measure its effectiveness.}

\subsubsection{Direct Assessment}
Direct assessment of counter-misinformation through quantitative questions involves various research methodologies, including in-lab experiments, interviews, and surveys.\footnote{Some reviewed papers use the term questionnaire. In this work, we use ``survey'' to cover questionnaire- as well as survey-based methodologies.} 
\textbf{In-lab experiments} are a common approach in many studies \cite{Stojanov2015, Kim2021, pasquetto2022social, kessler2022debunking, tay2022comparison, dai2022effects, pillai2022does, Masullo2021, Vraga2021}. Crowds are presented with both misinformation and various types of countering-misinformation articles. They then answer questions to assess the plausibility and believability of misinformation or counter-misinformation \cite{yu2022correcting}. 
As an advanced design for this methodology, \citet{Orosz2016} \old{ask crowds to first evaluate their initial belief in the misinformation} \new{ask crowds their initial belief in misinformation}, \old{view the counter-misinformation} {present them counter-misinformation}, and \old{then} \new{finally} reevaluate their belief. 
The change in belief \old{serves as a measure of} \new{measures} the effectiveness of counter-misinformation. 
\old{
In-lab experiments investigate diverse countering strategies, including the impact of news sources, both mainstream and private \cite{Mancosu2021}, and other factors \cite{Meer2020, Tanaka2019, Kim2021, Allen2021, Featherstone2020, Pennycook2019, Soprano2021, Kirchner2020, Bode2018, McCreadie2015, Martel2021, Sun2021}.
}
\new{
These in-lab experiments investigate diverse factors in countering strategies~\cite{Meer2020, Tanaka2019, Kim2021, Allen2021, Featherstone2020, Pennycook2019, Soprano2021, Kirchner2020, Bode2018, McCreadie2015, Martel2021, Sun2021}, e.g., independent or mainstream news sources~\cite{Mancosu2021}.
}

\begin{table}[t]
    \centering
    \caption{Taxonomy of crowd evaluators for counter-misinformation effectiveness.}
    \label{tab:Evaluation Metrics and Methods}
    \begin{tabular}{llll} \toprule
       & \multicolumn{2}{c}{\textbf{Categories}}   & \textbf{References} \\ \midrule
      \multirow{6}{*}{\rotatebox{90}{\textbf{(a) Methods}}} & \multirow{3}{*}{Direct}  & In-lab Experiment  &  \cite{Stojanov2015, Kim2021, pasquetto2022social, kessler2022debunking, tay2022comparison, dai2022effects, pillai2022does, Masullo2021, Vraga2021, yu2022correcting, Meer2020, Tanaka2019, Kim2021, Allen2021,  Featherstone2020, Pennycook2019, Soprano2021, Kirchner2020, Bode2018, McCreadie2015, Martel2021, Sun2021} \\ 
      & & Interview  &   \cite{Borah2021} \\   
      & & Survey  &  \cite{kessler2022debunking, Kirchner2020}  \\  
      \cmidrule(lr){2-4}
      & \multirow{3}{*}{Indirect} & Sentiment  &  \cite{zhang2022investigation}  \\  
      & & Stance  &  \cite{wang2022factors, wang2021evaluating} \\ 
      & & Number of Shares  &    \cite{chen2021citizens} \\ \midrule
      \multirow{6}{*}{\rotatebox{90}{\textbf{(b) \textbf{Findings}}}} & \multicolumn{2}{c}{Media Formats} & \cite{Kirchner2020, Veeriah2021, yu2022correcting, yang2022if, Vraga2021, kessler2022debunking, Masullo2021, pasquetto2022social} \\
      & \multicolumn{2}{c}{Content Traits} &  \cite{wang2021evaluating, kessler2022debunking, Orosz2016, Stojanov2015, Kirchner2020, yu2022correcting}  \\
      & \multicolumn{2}{c}{Communication Styles} & \cite{Kim2021, Masullo2021, Orosz2016, wang2022factors, yu2022correcting} \\
      & \multicolumn{2}{c}{Audience Factors} &  \cite{tanihara2022effects, wang2022factors, pasquetto2022social, yang2022if, primig2022influence} \\
      & \multicolumn{2}{c}{Countering Order} &  \cite{dai2022effects, tay2022comparison} \\
      & \multicolumn{2}{c}{Word Placement} &  \cite{pillai2022does} \\
      \bottomrule       
    \end{tabular}
    
\end{table}

\textbf{Interviews} offer personal interactions for crowds to engage in conversations and \old{ask} \new{answer} open-ended questions about effective strategies for countering misinformation.
During the COVID-19 pandemic in 2020, \citet{Borah2021} interviewed young adults and active social media users regarding their COVID-19 perceptions, coping strategies, and recommended countermeasures. 
\old{Findings advocate calling out people sharing misinformation and bringing up media literacy programs to combat misinformation.}
\new{These young crowds recommend calling out people who share misinformation and bring up media literacy programs.}

Finally, online \textbf{surveys} efficiently expand sample sizes and enhance inclusivity and accessibility for geographically dispersed \old{or busy} participants, thereby complementing in-lab experiments and interviews. 
These surveys collect demographic information and gather crowd opinions on misinformation and counter-misinformation, including how individuals handle potential false posts~\cite{Kirchner2020} and their evaluations of countering efforts~\cite{kessler2022debunking}.

\subsubsection{\textcolor{black}{Indirect Assessment}}

\new{
In contrast to direct assessment, indirect assessment utilizes proxy metrics to measure the effectiveness of counter-misinformation messages in a data-driven manner.  
For instance, \citet{zhang2022investigation} analyzes the \textbf{sentiment} in the comments on and replies to counter-misinformation messages by crowds on the Sina Weibo platform. 
They mention that positive sentiment is a valid indicator to model the efficacy of counter-misinformation messages. 
Practically, they use a Python library named SnowNLP to extract sentiment scores from the text~\cite{zhang2022investigation}. 
To have a comprehensive measurement, they also consider the number of likes and retweets in a weighted sum manner together to derive the final effectiveness score.
Similarly, other works assess the \textbf{stance} expressed in comments on rumor rebuttals as a proxy for countering acceptance \cite{wang2022factors, wang2021evaluating}. 
They first categorize stances as supporting, denying, questioning, and commenting and compute the stance of each comment by a BERT-based textual classifier~\cite{wang2021evaluating}. 
Next, for each rumor rebuttal, they use the ratio of ``the diﬀerence between the denying and supporting stances'' to ``the sum of the two stances, i.e., omitting other stances that do not contribute to the veracity of a rumor from all stances'' to calculate the degree of denial. This ratio could normalize the denying and supporting stances and make the metric comparable across rumors and events after dividing the redundant comments among all comments. 
Finally, the inverse of this normalized ratio indicates the debunking effectiveness~\cite{wang2021evaluating}. 
Additionally, \citet{chen2021citizens} use the \textbf{number of shares}. 
They particularly count the number of shares or reposts of fact-checks and explore the influence of peripheral cues (e.g., media richness and source credibility) and central cues (content importance and content theme) on the number of shares. 
Results show that the peripheral and central cues play critical roles in the sharing of fact-checks. 
For example, media richness and content importance significantly promote the number of shares.
}

\subsection{Findings from Effective Counter-Misinformation}
\label{sec:5.2 crowd evaluator findings}

After establishing the appropriate crowd-based evaluation methods, we utilize \new{these} crowd evaluators to gain \old{comprehensive} insights into the effectiveness of counter-misinformation efforts. 
Specifically, we aim to uncover the underlying factors contributing to these endeavors, \old{considering dimensions} such as media formats, content traits, communication styles,  audience factors, countering order, and word placement.

\subsubsection{Media Formats}
We examine the impacts of different formats employed in counter-misinformation efforts. 
\old{The predominant format is to respond to textual misinformation with text-based countermeasures \cite{Kirchner2020, Veeriah2021, yu2022correcting, yang2022if}.}
\new{
The predominant format is text-based countermeasures responding to textual misinformation~\cite{Kirchner2020, Veeriah2021, yu2022correcting, yang2022if}.
} 
Furthermore, a study by \citet{Vraga2021} explored text-based responses to video misinformation on health topics. 
Crowd evaluators viewed misinformation videos and accompanied debunking comments. The result showed that real-time crowd debunking in text partially reduced belief in misinformation ($p < 0.01$). 
\old{Additionally, \citet{kessler2022debunking} investigated text and images, finding credible text-based debunking supported by evidence effectively countered misinformation, while images alone had limited impact. }
\new{
Additionally, \citet{kessler2022debunking} investigated the format of text plus images. 
They found that credible text-based debunking supported by evidence effectively countered misinformation while images alone had limited impact. 
}
\old{Moreover, \citet{Masullo2021} studied uncivil countering comments with emojis, discovering that while the uncivil countering text has effects on the perception of the credibility of the news stories, reactions like ``angry'' can mitigate incivility and reduce dislike of news or comments, especially across different political leanings ($p < 0.01$). This finding potentially indicates the efficacy of emojis when countering misinformation. } 
\new{
Moreover, \citet{Masullo2021} studied the countering text that is accompanied by emojis. 
They discovered that while the uncivil expression impedes the perceived credibility of news stories, emoji reactions like ``angry'' mitigate incivility, especially across different political leanings ($p < 0.01$). 
}
Meanwhile, \citet{pasquetto2022social} found that audio-based corrections on WhatsApp platforms generate more interest and are more effective than text- or image-based messages in countering misinformation ($p=0.01$).

\subsubsection{Content Traits}

We further delve into the \old{key} content characteristics that are crucial to \old{effectively countering} \new{counter} misinformation. 
\old{These include the ability to uncover the motivations behind misinformation~\cite{Stojanov2015}, the use of evidence-based and logical counter-arguments~\cite{wang2021evaluating, Orosz2016}, the inclusion of warnings to alert readers to potential misinformation~\cite{Kirchner2020}, and the careful selection of credible sources for correction~\cite{yu2022correcting}, which are elaborated on below. }
\new{
These include uncovering the motivations behind misinformation~\cite{Stojanov2015}, using evidenced and logical counter-arguments~\cite{wang2021evaluating, Orosz2016}, containing warnings to remind readers of potential misinformation~\cite{Kirchner2020}, and selecting credible sources for correction~\cite{yu2022correcting}.
We elaborate on them below. 
}

Regarding \old{revealing misinformation motives} \new{uncovering the motivations behind misinformation}, 
\citet{Stojanov2015} examined the impact of debunking messages that reveal the motives behind conspiracy theories, particularly in the context of \new{vaccine-related medical topics}. 
Their findings suggest that \old{revealing the motives of conspiracy theorists} \new{this revelation} \old{can effectively reduce} \new{indeed reduces} belief in medical conspiracy theories ($p < 0.05$), although the effect on \old{belief in} general conspiracy theories is less clear.
\new{For using evidenced counter-arguments},
\citet{wang2021evaluating} highlighted the importance of citing evidence to enhance the debunking effect, which is better than debunking \old{methods} with uncited evidence ($p<0.001$).
\new{Subsequently, \citet{kessler2022debunking} extended these findings by confirming that the text with evidence rather than images matters.}
\new{
To investigate logical counter-arguments, \citet{Orosz2016} presented them with conspiracy theory statements to subjects and found that this approach reduces belief in conspiracy theories ($p \le 0.01$). 
}
\new{
\citet{Stojanov2015} additionally demonstrated that the countering works when it points out the fallacies in the reasoning behind the original misinformation post.
}
\new{
Concerning \textbf{displaying warning messages},
\citet{Kirchner2020} revealed that German adults preferred them as a strategy to combat fake news (65\% agreement compared to the baseline of 51-57\%). 
They also highlighted providing explanations for flagged misinformation (71\% agreement compared to the baseline of 51-57\%).
These results demonstrate the effectiveness of warning-based countering, particularly when accompanied by explanatory text.
}
\new{
Lastly, about \textbf{selecting credible sources for correction}, researchers identified variations in the persuasiveness of different sources~\cite{yu2022correcting}.
For example, \citet{yu2022correcting} found that in authoritative societies, corrections from government sources tend to have greater credibility than those from professionals or laypeople ($p \le 0.05$). 
This underscores the need to tailor source selection to specific societal contexts, particularly when the impact of corrections varies across different societies (e.g., authoritative societies). 
}

\subsubsection{Communication Styles}

Analyzing different communication styles of counter-misinformation, such as humorous and uncivil responses, provides an additional dimension to evaluating its effectiveness. 
\new{
Regarding \textbf{humor}, researchers have explored its potential by comparing humorous and non-humorous corrective messages to HPV vaccine-related misinformation~\cite{Kim2021}.
} 
\new{
They found that humor increased the visual attention of crowds toward the image portion of the counter-misinformation counts. 
This indirectly reduced HPV misperceptions. 
} 
\new{For} \textbf{uncivil responses}, \citet{Masullo2021} investigated 
\old{the effect of uncivil counter-misinformation comments} 
\new{its effect}
on readers' attitudes and perceptions. 
In an online survey, crowd evaluators were exposed to misinformation articles, uncivil comments, and corresponding social reactions to the comments, followed by questions about their feelings. 
The findings revealed that while uncivil counter-comments influenced how readers viewed the comments and the commentators themselves, they did not affect the credibility of the underlying articles. This suggests that uncivil responses may not contribute significantly to countering misinformation.

In addition, the literature has explored other factors such as \old{\textbf{ridiculing vs. empathetic}} \new{\textbf{ridicule}, \textbf{empathy}}, \textbf{readability}, and  \textbf{tone of correction}. 
According to \new{research works by}  \citet{Orosz2016}, ridiculing arguments \old{effectively} reduce belief in conspiracy theories, while empathetic counterarguments have no significant impact \old{on belief in misinformation}.
\citet{wang2022factors} investigated the \old{impact} \new{effect} of the readability \old{of the rebuttal text on crowds' acceptance of the rebuttal. }  
\old{The study} \new{They} found that higher readability positively influenced users' acceptance of the rebuttal. 
Improved readability indicates that users can easily understand and absorb the rebuttal, thereby enhancing its effectiveness.
Lastly, \citet{yu2022correcting} examined the impact of the tone of corrective messages \old{on the believability of the correction}. 
\old{The} Their research indicated that a formal tone was more believable than a less formal and conversational tone.

\subsubsection{Audience  Factors}
   
In addition to the inherent characteristics of counter-misinformation, a number of external factors related to the audience, including media literacy, cognitive capability, political stance, trust in media, and concern about misinformation, also \old{have an impact on} \new{influence} its effectiveness.

Research has extensively explored the impact of  \textbf{media literacy} on the perception of misinformation. 
For instance, \citet{tanihara2022effects} found that crowd evaluators with lower levels of media literacy are more likely to change their perceptions \old{of misinformation} when exposed to corrective messages. 
In a study conducted by \citet{wang2022factors}, 
\old{
\textbf{cognitive capability} was assessed based on an individual's language-related cues, including vocabulary size. The findings revealed that
}
\new{
they assessed \textbf{cognitive capability} based on an individual's language-related cues (e.g., vocabulary size) and found that
}
crowd evaluators with lower cognitive capabilities tend to rely on the credibility of the source and the quality of the argument in the rebuttal text to accept it. 
On the other hand, \old{readers} \new{crowd evaluators} with high cognitive capability have a \old{greater} \new{higher} demand for \old{greater} readability, even when the sources and arguments are solid.  
\old{Regarding} For \textbf{political stance}, \citet{pasquetto2022social} found that the correction from people with strong ties (e.g., the same political stance) can lead crowds to more actively re-share these debunks and counter misinformation. 
This tendency was corroborated by \citet{yang2022if}, who observed that crowd evaluators are more likely to accept corrections from sources that align with their existing attitudes. 
\old{Regarding} \new{Concerning} \textbf{trust in media},  
\citet{primig2022influence} discovered that higher trust in media sources corresponds to higher believability in fact-checking sources and messages, especially when combined with \new{user} trust in politics. 
\old{
Conversely, for social corrections on social media platforms,  their effectiveness is more pronounced among users who experience higher levels of uncertainty on these platforms.
}
\new{
Conversely, \citet{yang2022if} noted that lower levels of certainty on social media platforms lead to less effectiveness of corrections on these platforms.
}
\old{Regarding} \new{For} \textbf{concern about misinformation}, \citet{yang2022if} found that social corrections work better among users who are more concerned or worried about the potential harm caused by misinformation on social media platforms.

\subsubsection{Countering Order and Word Placement} 

Researchers have also examined other less prominent factors such as the order of counter-misinformation and the placement of negation.
Regarding \textbf{countering order}, two contrasting approaches emerge: debunking starts by presenting misinformation and then offers a counterresponse, while prebunking reverses this order by introducing counter-misinformation as a preemptive warning or reminder. 
\citet{dai2022effects} confirmed that prebunking messages, especially when combined with fact-checks or inoculation messages, \old{effectively} made crowd evaluators more skeptical of misinformation.
\old{Additionally, } \citet{tay2022comparison} conducted a more comprehensive evaluation that extended \old{beyond} traditional questionnaires to assess user behaviors \old{, such as information seeking and the promotion of online misinformation} \new{(e.g., information seeking and misinformation promoting)}.  
The results highlighted the effectiveness of both prebunking and debunking in reducing belief in misinformation. 

Another \new{studied} nuanced aspect \old{studied} is how the \textbf{placement of negation} during debunking impacts crowd memory and, consequently, the effectiveness of debunking~\cite{pillai2022does}.
\old{In} \new{On} social media, brief affirmations or negations are often used to clarify claims. 
In this context, \citet{pillai2022does} found that when crowd evaluators encounter negated messages (\eg, ``This is wrong.''), they often remember the main claim but forget the negative part. 
They tested placing the negation before or after the entire claim and found that both methods were equally memorable. 
\old{This result highlights} \new{This finding underlines} the possibility of diverse approaches to conveying affirmations and negations \new{for misinformation correction}.

\section{Crowds as Creators of Counter-Misinformation}\label{sec:analyze_the_spread_of_existing_counter}

\begin{figure}[t]
\centering
\includegraphics[width=0.98\linewidth]{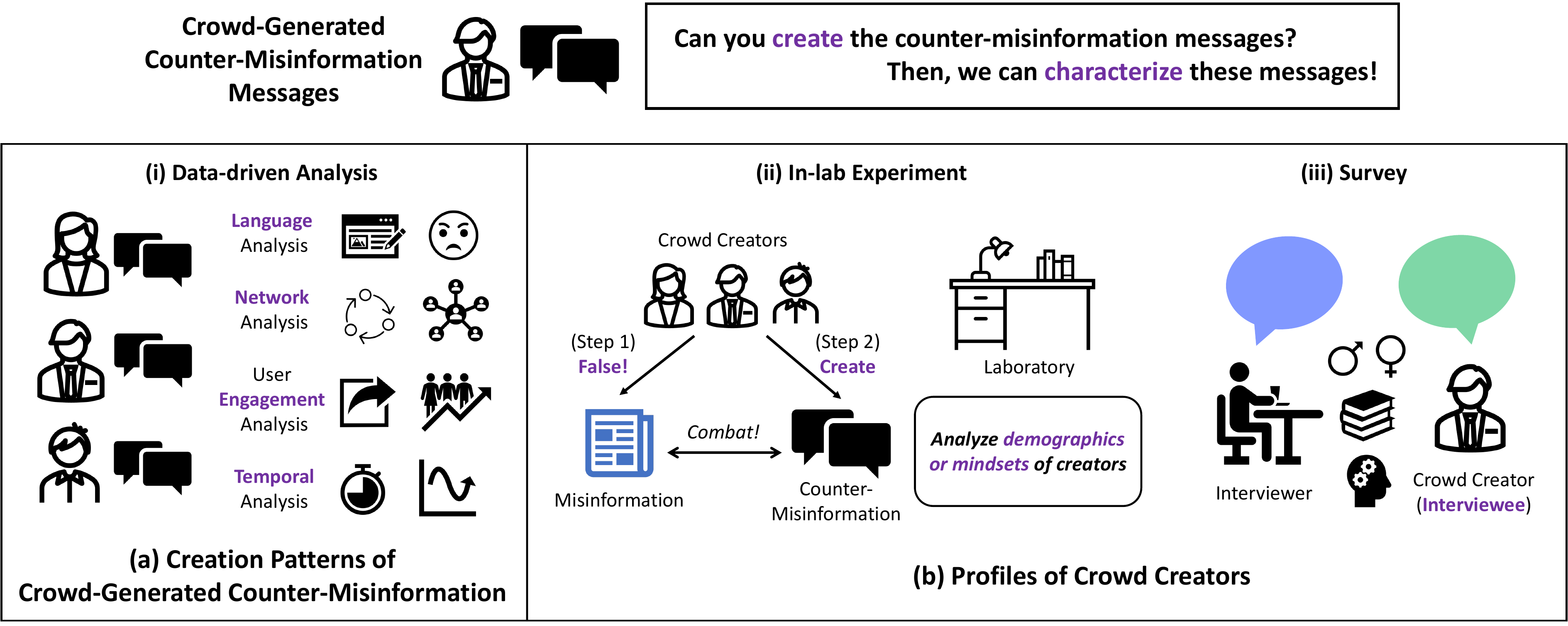} 
\caption{Illustration of crowds' role as creators of counter-misinformation.} 
\label{fig:creators}
\end{figure}

In addition to serving as annotators and evaluators, crowds \old{have the ability to} \new{can} create their own counter-misinformation messages on social media platforms.
In this section, we aim to examine their role as content creators (see Figure~\ref{fig:creators}) to investigate their unique perspective and proactive involvement in combating misinformation while characterizing their content. 
To achieve this goal, we first present three characterization methods that measure how crowds actively combat misinformation as creators.
These methods are data-driven analysis, in-lab experiment, and survey (see Table~\ref{tab:findings of creator}-(a)).
Based on these methods, we discuss findings \old{on} \new{after} analyzing \new{crowd-generated} counter-misinformation \old{generated by crowds}, with a \old{specific} focus on creation patterns \new{of counter-misinformation} and \new{its} creator profiles (see Tables~\ref{tab:findings of creator}-(b) and (c)).
\old{
Specifically, we conduct a comprehensive examination of counter-misinformation generated by crowds, revealing the creation patterns of counter-misinformation.
Finally, we explore the intricate dynamics of the human elements involved, focusing on the profiles of crowd creators. 
} 

\begin{table}[t]
    \centering
    \caption{Taxonomy of crowd creators for analyzing crowd-generated counter-misinformation.}
    \label{tab:findings of creator}
    
    \begin{tabular}{llll} 
        \toprule
        & & \multicolumn{1}{l}{\textbf{Categories}}  & \textbf{References} \\ 
        \midrule
        \multirow{3}{*}{\rotatebox{90}{\textbf{(a) Methods}}} & 
        \multirow{2}{*}{\begin{tabular}{@{}c@{}}Creation\\ Patterns\end{tabular}} &  \multirow{2}{*}{\begin{tabular}{@{}l@{}}Data-driven  Analysis\end{tabular}}   &  \multirow{2}{*}{\cite{Mohammad2021, Micallef2020, xue2022covid, kim2022use, Hunt2020, Pal2017,giachanou2022impact}}  \\
        &&& \\ \cmidrule(lr){2-4}
      & \multirow{2}{*}{\begin{tabular}{@{}c@{}}Creator\\Profiles\end{tabular}}  & In-lab Experiment  &  \cite{Pal2018, Pal2019, Cohen2020, Igbinovia2021, Sun2020, huber2022taking, Veeriah2021} \\
      & & Survey  &  \cite{chen2022let}  \\  \midrule
        \multirow{13}{*}{\rotatebox{90}{\textbf{(b) Creation Patterns}}}

        & \multirow{6}{*}{Language}  & Sentiment & \cite{Mohammad2021, Micallef2020} \\  
        & & Emotion & \cite{xue2022covid, Micallef2020, kim2022use} \\ 
        & & Psycholinguistic Features & \cite{Micallef2020, xue2022covid, giachanou2022impact} \\ 
        & & URL & \cite{Micallef2020, Pal2017, Hunt2020} \\ 
        & & Hashtag & \cite{Gunaratne2019, Babcock2019JDIQ} \\ 
        & & Others & \cite{Zubiaga2016, Nadamoto2013, chen2022science}  \\ 
        \cmidrule(lr){2-4}
        
         & \multirow{3}{*}{Network}   & Network Co-existence Pattern & \cite{Gunaratne2019} \\ 
        & & Information Transmission & \cite{Zubiaga2016} \\ 
        & & User Connectivity Pattern & \cite{Ahmed2020,chiu2022election} \\ \cmidrule(lr){2-4}
        
        & \multirow{2}{*}{\begin{tabular}[c]{@{}l@{}}User\\Engagement\end{tabular}}  & Post Volume & \cite{wang2022exploring, Micallef2020} \\ 
        & & Post Sharing & \cite{Micallef2020, Chua2017} \\ 
        \cmidrule(lr){2-4}        
         
         &  \multirow{2}{*}{Temporal} & Temporal Trend & \cite{Gunaratne2019} \\ 
        & & Life Cycle & \cite{Zubiaga2016} \\ 
        \midrule

      \multirow{8}{*}{\rotatebox{90}{\textbf{(c) Creator Profiles}}} & \multirow{3}{*}{Demographics}  & Age and Education  & \cite{vijaykumar2022dynamics} \\ 
      && Political Ideology & \cite{Cohen2020, steinfeld2022disinformation} \\ 
      && Media Literacy & \cite{Igbinovia2021, vijaykumar2022dynamics, Veeriah2021, huber2022taking} \\
    \cmidrule(lr){2-4}
        
      & \multirow{4}{*}{Mindsets} 
      & Sharing Denials & \cite{Pal2018, Pal2019} \\
      && Others' Susceptibility & \cite{Sun2020} \\ 
      && Third-person Effect & \cite{chen2022let} \\ 
      && Harm Awareness & \cite{Zhao2016, vijaykumar2022dynamics} \\
    % \cmidrule(lr){2-4}         
      \bottomrule
      
    \end{tabular}
\end{table}

\subsection{Characterization Methods}
\label{sec:Characterization Methods}

To establish a \old{comprehensive} framework for analyzing the \old{active} counter-misinformation creation process~\old{involving human elements}, we present three characterization methods in Table \ref{tab:findings of creator}-(a). 
% \bing{the following sentence may not be relevant in the context}
\old{We note that different methods have been utilized depending on what we aim to characterize.}
First, researchers mainly rely on \textit{data-driven analysis} to understand the \textbf{creation patterns} of effective crowd-generated counter-misinformation. 
\old{In a nutshell} \new{In essence}, they \old{begin by crawling} \new{first crawl} large-scale crowd-generated data, including posts and social network relationships. 
\old{This data is then examined from different perspectives} \new{They then examine different perspectives in the data}, such as language patterns, networks, and temporal trends  \cite{Mohammad2021, Micallef2020, xue2022covid, kim2022use, Hunt2020, Pal2017,giachanou2022impact}, which will be elaborated in Section~\ref{sec:creation patterns}. 

On the other hand, researchers employ a combination of \textit{in-lab experiments and surveys} to study \textbf{creator profiles}.
% in Section \ref{sec:creator profiles}, 
For in-lab experiments  \cite{Pal2018, Pal2019, Cohen2020, Igbinovia2021, Sun2020, huber2022taking, Veeriah2021}, crowd creators are exposed to misinformation \old{text} and then asked \old{to respond by either countering or endorsing it} \new{whether to counter or endorse it}. 
In the meantime, researchers collect information \old{on} \new{about} creators, including their political leanings and beliefs related to misinformation, \old{in order } to analyze their attributes. 
Additionally, researchers conduct online surveys to assess the creators' levels of belief in misinformation and collect demographic information, \old{including} \new{e.g., } their media competency  \cite{chen2022let}. 
The results are discussed in Section~\ref{sec:creator profiles}.

The goal of these methods is to characterize the creators themselves and uncover the patterns and attributes associated with crowd-generated counter-misinformation.
In this context, it is important to note that this objective contrasts with the goal of the crowd-based evaluation methods~\new{in Section~\ref{sec: crowds as evaluators}}, which assess misinformation countermeasure effectiveness.

\subsection{Creation Patterns}
\label{sec:creation patterns}

In the battle against misinformation, crowds actively contribute by generating a multitude of \new{counter-misinformation} content \old{that includes} \new{e.g., } debunking posts and responses. 
\old{A significant amount of} \new{Extensive} research has \old{directed its attention towards} \new{focused on} comprehending how such counter-misinformation messages are crafted and pinpointing the patterns they contain \cite{Micallef2020, Chua2017, Nadamoto2013, kim2022use}. Researchers typically first manually label sampled text and visuals into categories such as misinformation and counter-misinformation, and then train classifiers \old{for extensive categorization on} \new{to categorize} unlabeled data points due to the high cost of manual labeling.
This \old{comprehensive} categorization serves as the basis for analyzing counter-misinformation, including content characteristics, network dynamics, temporal aspects, and social media engagement.  For a detailed breakdown of these dimensions, refer to Table \ref{tab:findings of creator}-(b).

\subsubsection{Language Analysis}

Crowd-generated counter-misinformation messages exhibit distinct language patterns when compared to other online content \cite{Mohammad2021, xue2022covid, kim2022use}. We summarize these patterns through various aspects, including sentiments, emotions, psycholinguistic features, and textual components like URLs and hashtags.

First, \textbf{sentiment} analysis revealed differences among various topics.  
For instance, during the COVID-19 pandemic, positive attitudes were expressed in comments countering COVID-19 conspiracy theories on Facebook \cite{Mohammad2021}, while counter-misinformation tweets regarding fake COVID-19 cures were more neutral \cite{Micallef2020}. 
In addition, \textbf{emotions} \old{expressed in counter-misinformation} have been explored in various contexts, demonstrating that contexts affect different emotional responses. 
For \old{instance} \new{example}, \citet{xue2022covid} used IBM Watson Tone Analyzer to extract emotions, revealing that COVID-19 vaccine fact-checking posts on Facebook tended to maintain a neutral tone, while public comments often displayed more emotionally charged responses. 
\old{Additionally, } \citet{kim2022use} \new{further} analyzed emotions in videos related to COVID-19 on YouTube.
They utilized the NRC Emotion Lexicon\footnote{\url{http://saifmohammad.com/WebPages/NRC-Emotion-Lexicon.htm}}, 
which contains a list of English words and their associations with emotions and sentiments.
The study identified two emotional dimensions, trust and fear, and observed differential utilization of these emotions in debunking videos.

Furthermore, researchers have delved into the \textbf{psycholinguistic features} \old{of counter-misinformation} using tools such as Linguistic Inquiry and Word Count (LIWC) or IBM Watson Tone Analyzer \cite{Micallef2020, xue2022covid, giachanou2022impact}.
For instance, on Facebook, posts fact-checking COVID-19 vaccine information exhibited higher levels of confidence compared to general public posts \cite{xue2022covid}. 
Additionally, \citet{giachanou2022impact} discovered that crowds tend to employ more positive \old{language} and causal language, whereas misinformation spreaders often use more informal language.

\old{Additionally} Two \old{noteworthy} elements frequently integrated into counter-misinformation messages are URLs and hashtags.  
Uniform Resource Locators (\textbf{URLs}) serve as sources of evidence-based resources  \cite{Micallef2020}, often cited in tweets that counter rumors \cite{Pal2017}. 
Moreover, URLs have proven effective in disseminating information for debunking during disaster events, often citing news agencies as their primary information sources \cite{Hunt2020}.
\textbf{Hashtags}, on the other hand, are employed to enhance the visibility and categorization of content. For example, \citet{Gunaratne2019} analyzed vaccine-related tweets and found that 86\% of users exclusively used pro-vaccine hashtags,  while 12\% opted for anti-vaccine hashtags. Additionally, \citet{Babcock2019JDIQ} found that the hashtag ``\texttt{\#fakenews}'' was widely used to attack fake news.

Besides these specific content attributes, \textbf{other dimensions} like stance \cite{Zubiaga2016}, user impression \cite{Nadamoto2013}, politeness~\cite{Micallef2020}, and the group identity language~\cite{chen2022science} have also been examined in counter-misinformation.

\subsubsection{Network Analysis}

Researchers employ various network analysis approaches and tools (e.g., NodeXL and Gephi~\cite{Flores2021, Jung2020, Babcock2019, Memon2020, Dang2016, Weber2020}) to uncover the intricate interplay between misinformation/counter-misinformation information flow and user connections on social media platforms. The representative approaches encompass patterns of co-existence in networks, post sharing/(re)tweeting, and connectivity of users.

Regarding \textbf{network co-existence patterns}, \citet{Gunaratne2019} examined the co-occurrence of anti-vaccine and pro-vaccine hashtags on Twitter. 
\old{Their analysis involved constructing} \new{They constructed} a network with hashtags as nodes and co-occurring hashtags as edges, revealing distinct community structures within \old{both} pro-vaccine and anti-vaccine hashtags. 
The results highlighted that pro-vaccine hashtags formed a dominant community along with a few closely connected sub-communities.
Conversely, anti-vaccine hashtags largely converged into one community with a smaller, remote sub-community that focuses on specific vaccines. 
This analysis enhanced our understanding of the discourse and inter-relationship between misinformation and counter-misinformation.

To investigate \textbf{information transmission} within a social network,  \citet{Zubiaga2016} investigated the rumor (re)tweeting networks which included three types of retweets:
(1) unverified source tweets, (2) accurate tweets supporting true information or denying false rumors (\ie, counter-misinformation), and (3) inaccurate tweets denying true information or supporting false rumors (\ie, misinformation). 
By examining retweeting relationships between (re)tweet nodes, they discovered that retweets of accurate tweets were primarily observed in certain misinformation topics, such as the Ottawa shooting and the Sydney siege. In contrast, tweets sharing unverified rumors had a broader spread across different topics.

\textbf{User connectivity patterns} on social network platforms such as Twitter have also been studied in the literature \cite{Ahmed2020,chiu2022election}. 
For instance, during the COVID-19 pandemic on Twitter, \citet{Ahmed2020} regarded crowds as nodes and the ``\texttt{reply-to}'' or ``\texttt{mention}'' relationships in tweets as edges in a graph. 
A ``\texttt{self-loop}'' relationship is added in the graph if there is no ``\texttt{reply-to}'' or ``\texttt{mention}'' in the tweets.
By analyzing such a graph, researchers found that $32.2\%$ of the crowds denounced the COVID-19 5G conspiracy theory.
They also identified the two largest network structures, consisting of an isolated group and a broadcast group. 
The results revealed that there was \old{ a lack of an} \new{no} active authority figure to counter the spread of misinformation. 
Instead, the crowds countered the conspiracy theory with widespread denouncement.
Similarly,  \citet{chiu2022election} considered the ``\texttt{mention}'',  ``\texttt{retweet}'', and ``\texttt{self-loop}'' relationships between users to create the graph and conduct the cluster analysis, while focusing on the diffusion scope, speed, and shape for true and fake news across the users. 
They also tested whether the attributes of the true or fake news spreaders would affect the aforementioned three metrics. 
The results showed that true news from the crowd tended to spread later and with less broadcast influence compared to fake news.
In addition, \citet{wang2022exploring} investigated the follower-followee relationships of refuters and non-refuters. They developed a deep-learning-based text classifier to identify debunked and non-debunked posts on Sina Weibo. They then identified associated refuters and non-refuters whose follower-followee networks are crawled as well. Their analysis unveiled that nodes with greater centrality had more follower-followee edges, and weakly connected components could easily disseminate both debunked and non-debunked posts. This suggests that misinformation and counter-misinformation have similar propagation patterns, considering that both can spread from \old{weekly-connected} \new{weakly-connected} nodes.

\subsubsection{User Engagement Analysis}

Social media engagement analysis explores how crowds interact with online posts containing both misinformation and counter-misinformation. 
\old{
These interactions can be measured through metrics like post volume growth, and sharing behaviors can differ depending on specific scenarios.
} 
\new{
These interactions can be measured through metrics like post volume growth, and post sharing behaviors. 
}

For example, \citet{wang2022exploring} observed that the \textbf{post volumes} of counter-misinformation exhibited growth rates similar to those containing misinformation.
Nonetheless, \citet{Micallef2020} uncovered disparities in their absolute quantities and growth trends during the COVID-19 pandemic. 
Specifically, the number of counter-misinformation posts was significantly lower than the number of misinformation posts. 
Apart from post volumes, they also analyzed the imbalanced \textbf{sharing behavior} associated with these posts, and found that the majority of posts, regardless of their classification as misinformation or counter-misinformation, received minimal to no shares. 
However, specific situations could lead to variations. 
For example, when examining the rumored death of Singapore's President Lee Kuan Yew, \citet{Chua2017} found that tweets aiming to correct the rumor garnered more frequent retweets than the initial rumor tweets. This suggests that, in certain contexts, counter-misinformation may be more prone to be shared than misinformation.

\subsubsection{Temporal Analysis}

Temporal analysis aims to uncover evolving misinformation and counter-misinformation traits by studying temporal trends and life cycle patterns, offering valuable insights into their dynamics.

For instance, \citet{Gunaratne2019} examined the \textbf{temporal trends} of disease cases, pro-vaccine of diseases, and anti-vaccine of diseases tweets, as well as the crowds who tweeted between 2010 and 2019. 
The results revealed that pro-vaccine tweets consistently outpaced anti-vaccine tweets in volume, and this trend continued to increase over time.
Similarly, \citet{Zubiaga2016} conducted a study on the \textbf{life cycle} of rumors and their countering. 
They aimed to understand how users engage with rumors, in terms of both their spread and debunking, before and after the veracity of the rumor is confirmed. 
Through manual \old{identification and} annotation of rumor threads, \new{they had} interesting observations \old{were made}. 
\old{Notably} \new{For example}, rumors that were eventually proved to be true, were debunked more quickly than false rumors. 
\old{Interestingly,} When a rumor tweet was countered by either the crowd or an organization, retweets occurred more evenly over time, indicating sustained retweeting activities. 
These findings \old{shed light on} \new{illustrated} the dynamics of rumor propagation and debunking activities over time.

\subsection{Creator Profiles}
\label{sec:creator profiles}

Beyond characterizing the creation patterns of crowd-generated counter-misinformation, we also analyze the profiles of these crowds as counter-misinformation creators. 
Specifically, we examine their demographic factors and mindsets, as outlined in Table \ref{tab:findings of creator}-(c).

\subsubsection{Demographics}

Literature highlights the \old{significant} influence of certain demographic factors on crowds' willingness and involvement in countering misinformation. 
One noteworthy aspect is \textbf{political ideology}, which influences one's approach to governance and public policy. 
According to \citet{Cohen2020}, the \old{presence of a} social identity threat in fake news \old{content} related to one's political ingroup indirectly affects crowds' readiness to publicly denounce fake news \old{articles}, \new{finally attacking the political ideology}. 
Similarly, \citet{steinfeld2022disinformation} found that individuals who engage in violent or illegal political protests tend to actively combat disinformation, even though they may occasionally share disinformation themselves. 
These findings underscore the intricate interplay between political ideology and countering misinformation.

\textbf{Media literacy}, also known as ``Information Literacy Competence'' and ``News Media Literacy,'' is another widely investigated factor. 
It involves the critical analysis of media content, particularly news, to discern its credibility, bias, and intention.
Several studies have highlighted its absence among the general audience, as well as the necessity for improved media literacy education. 
For instance, \citet{Igbinovia2021} emphasized the importance of media literacy in limiting the spread of fake news during the COVID-19 pandemic.
They revealed that greater media literacy can aid in identifying fake news, which in turn may facilitate \old{social correction} \new{misinformation combating} ~\cite{vijaykumar2022dynamics}. 
\citet{Veeriah2021} \new{also} affirmed the crucial role of media literacy in motivating corrective responses to misinformation. 
However, they found that even young crowds who felt sure about identifying fake news still demonstrated only moderate levels of media literacy. This observation implies the necessity of initiating media literacy education amongst young crowds \old{to bolster the fight against fake news}.
Similarly, \citet{huber2022taking} echoed this finding, indicating that relying solely on general media literacy is insufficient in \old{effectively} countering certain types of fake news.  
These collective findings highlight the significance of high-level or even specialized media literacy for crowds engaged in countering misinformation.

In addition to political ideology and media literacy, the above studies also examined additional factors such as \textbf{age} and \textbf{education}. 
For instance, \citet{vijaykumar2022dynamics} found that crowds who are younger or less educated tend to be less involved in countering misinformation.

\subsubsection{Mindsets}

The motivations of crowds to counter misinformation are influenced by their beliefs and mindsets.  
For instance, \old{the act of} \textbf{sharing denials} has been identified as a crucial strategy for countering misinformation~\cite{Pal2018, Pal2019}.
\citet{Pal2018, Pal2019} \old{delved into this area and} identified three key beliefs that encourage crowds to \old{share messages that deny rumors} \new{share denials}:  
(1) the belief that sharing denials helps spread the truth,
(2) the belief that friends and the online community would favor the behavior of sharing rumor denials, 
and (3) the belief that the credibility of the source of rumor denials encourages sharing, thus influencing crowds' willingness to share for countering. 

Moreover, we note \new{the \textbf{third-person effect} phenomenon} 
, denoting the belief among individuals that media messages have a greater influence on others than on themselves. 
For instance, \citet{chen2022let} found that this effect positively encourages crowds to debunk online misinformation. 
Another related belief, known as the ``\textbf{others' susceptibility}'', refers to the perception that ``others will be affected by or susceptible to misinformation''. 
\citet{Sun2020}  discovered that such belief \old{in others' susceptibility} can induce negative emotions such as guilt and anger, thus motivating vaccine supporters to correct anti-vaccination misinformation actively.

Lastly,  \textbf{harm awareness}, \old{or} \new{i.e., } being aware of the negative consequences, can also motivate efforts to counter misinformation.
For example, \citet{Zhao2016} found that perceiving the harm caused by rumors predicts a significant increase in engagement to counter misinformation. Similarly, \citet{vijaykumar2022dynamics} discovered that when crowds perceive that the severity of COVID-19 is not being properly addressed, they are motivated to \old{take action} \new{act} against related misinformation.
\section{{Discussion} and Future Directions}

\subsection{\new{Discussion}}

While our survey has focused on the potential of crowds~\cite{kumarcrowd} to combat misinformation through roles like annotators, evaluators, and creators, a nuanced comparative analysis of the effectiveness of each role is beneficial, especially highlighting which method should be employed under which circumstance: 
\begin{itemize}
    \item Annotators: annotating by crowds can be particularly effective for rapidly identifying, flagging, and fact-checking potential misinformation at scale (e.g., social media misinformation flagging, image and video verification). 
    The diversity of perspectives and cultural contexts within the crowd can aid in detecting nuanced forms of misinformation. 
    Practically, individual labeling is suited for (1) situations requiring unbiased, independent assessments; (2) initial stages of a study to establish a baseline of individual perceptions; (3) cases where privacy concerns prevent collaborative efforts; and (4) when trying to gauge the general public's ability to identify misinformation. 
    On the other hand, collaborative labeling is effective for (1) complex misinformation that requires diverse expertise to understand and evaluate; (2) fostering community engagement in combating misinformation; and (3) building consensus on challenging cases and developing shared standards for what constitutes misinformation and how to combat it. 
    Finally, machine-assisted labeling is useful for (1) large-scale misinformation detection efforts; (2) pre-screening content before human review; (3) augmenting human efforts in resource-constrained environments; and (4) rapidly evolving situations where quick identification is crucial.  
    However, for all these annotation methods, the quality and consistency of annotations may vary, necessitating robust quality control mechanisms and expert oversight. 
    \item Evaluators: Engaging crowds in evaluating counter-misinformation can provide valuable insights into the effectiveness of these interventions across diverse audiences (e.g., public health messaging and political advertising evaluation). The crowd's collective feedback can help refine and tailor counter-narratives for maximum impact. 
    Particularly, for direct assessment methods, we use in-lab experiments for (1) controlled studies of how people interact with and process counter-misinformation; and (2) testing the effectiveness of different counter-misinformation strategies. Interviews are appropriate for (1) in-depth exploration of individual experiences with counter-misinformation (e.g., understanding the personal and social contexts that influence belief in misinformation); and (2) gathering rich, qualitative data on how crows encounter counter-misinformation. Surveys are ideal for (1) collecting large-scale data across populations; and (2) tracking changes over time (e.g., gauging public awareness of specific misinformation and counter-misinformation campaigns).
    On the other hand, regarding indirect methods, 
    We analyze (1) sentiment for assessing emotional responses to counter-misinformation and tracking public mood around controversial topics prone to misinformation; (2) stance for assessing the alignment of information with established consensus in counter-misinformation and evaluating the effectiveness of counter-misinformation in changing stances; and (3) share/engagement analysis for measuring the reach and potential impact of misinformation and comparing the spread of misinformation versus counter-misinformation. 
    Nevertheless, evaluations may be susceptible to biases, and appropriate measures should be taken to ensure objectivity and representativeness.

    \item Creators: Leveraging the creativity and diverse perspectives of the crowd in generating counter-misinformation can lead to more contextually relevant and culturally resonant messaging (e.g., multilingual content creation). 
    In practice, we focus on the creation patterns of these crowd-generated counter-misinformation contents when analyzing and identifying the effectiveness of different counter-misinformation contents and assessing coordination with fact-checking from professional experts. 
    Besides, we emphasize the profiles of crowd creators when we (1) study the motivations and experiences of crowds, (2) identify influential counter-misinformation actors, and (3) assess diversity and representation in counter-misinformation, especially when we compare crowds with professional experts. 
    However, ensuring the accuracy and factual integrity of crowd-generated content may pose challenges, requiring rigorous fact-checking and editorial oversight.
    
\end{itemize}
As mentioned, each role's effectiveness varies based on task, context, and misinformation domain. 
For instance, annotation tasks may be more suitable for rapidly evolving narratives, while content creation may be better suited for addressing contextually relevant misinformation. 
By understanding these nuances, practitioners and policymakers can strategically employ crowds, leveraging the strengths of each role while mitigating potential limitations through quality control, expert oversight, and a combination of crowd- and expert-driven efforts.

\new{
To enhance the generalizability of crowd-based counter-misinformation efforts to real-world scenarios, it is essential to consider potential factors as follows:
\begin{itemize}
    \item Evolving misinformation tactics: Misinformation spreaders continuously adapt their tactics, potentially impacting the effectiveness of countermeasures by crowds. Continuous monitoring and adaptation are essential.
    \item Cultural and linguistic contexts: Misinformation narratives and their impact can be deeply rooted in specific cultural and linguistic contexts. Crowds need to take care of these nuances to accurately counter contextually-specific misinformation. 
    \item Regulatory and policy landscapes: Changes in regulations, policies, or legal frameworks governing online content and misinformation could impact the viability and implementation of crowd-based approaches, requiring adjustments to align with new guidelines or constraints.
\end{itemize}
By acknowledging these potential limitations in generalizability, our survey can offer more robust and actionable insights for real-world applications, contributing to the development of effective and sustainable crowd-based approaches to combating misinformation across diverse contexts.
}

\subsection{Future Directions}

In this section, we outline potential avenues for future research in this field:

\begin{itemize}

    \item \textbf{Improving Crowd Annotations:} While collaborative labeling aims to overcome the limitations in individual labeling (\eg,  individual annotation biases~\cite{Mitra2015}),  it's important to note that individuals often share similar perspectives (homophily). Therefore, it's unclear whether collaborative labeling effectively captures diverse viewpoints. 
    Future research should explore this aspect and develop strategies to encourage diversity within groups countering misinformation. Similarly, \old{crowd}\new{human}-in-the-loop identification of misinformation can benefit from agile classifiers that do not require extensive labeled examples initially  and can quickly  adapt to emerging misinformation \cite{mozes2023towards}. One approach to achieve this is by utilizing few-shot learning techniques~\cite{mozes2023towards}.

    \item \textbf{Multi-platform and Multimodal Countering:} Current social media-related work predominantly focuses on one specific platform like Twitter~\cite{he2021racism, Chua2017, Chua2017study}, Facebook~\cite{xue2022covid}, and Sina Weibo~\cite{wang2022factors}. However, crowds countering misinformation may behave differently across various platforms due to variations in user demographics and engagement dynamics. Exploring how crowds counter on multiple platforms and whether countering on one platform influences others is essential for a comprehensive understanding of crowd-driven misinformation mitigation. Additionally, the crowd-generated counter-misinformation is not limited to text alone; it can also involve \old{the use of} images or videos to enhance the persuasiveness of their debunking efforts.  Investigating  these multimodal aspects  benefits the design of effective countering contents.

    \item \textbf{Multilingual and Topic-specific Countering:} Most research works concentrate on either a single language (e.g., English~\cite{Chua2017study} or Chinese~\cite{wang2022factors}) or a specific misinformation topic (e.g., COVID-19~\cite{Veeriah2021}). But, misinformation spans across languages and topics, leading to diverse countering actions. Analyzing how crowds in under-representative languages combat various topics reveals variations in countering strategies across languages and topics, thus contributing to a comprehensive understanding of misinformation countering. On the other hand, existing research on the profiles of crowds focuses on demographic factors such as education, political leanings, and media literacy. However, it overlooks misinformation topic-specific factors. For instance, an individual having a background in health education may be effective at countering health misinformation but susceptible to believing in climate misinformation. Therefore, exploring these topic-specific factors can enhance our understanding of human factors involved in countering misinformation.

    \item \textbf{In-thread Countering:} While current research primarily examines the impact of standalone or accompanied counter-misinformation messages on social media platforms~\cite{Chua2017, xue2022covid},   there is limited exploration into the dynamics of counter-misinformation within conversations or threads~\cite{he2024corrective}. For example,  a crucial question arises: ``When others in a thread engage in countering, does it lead to correcting the misinformation or, conversely, amplifying it?'' 
    This form of in-thread countering (e.g., counter-misinformation responses to original misinformation posts) is frequently observed on social media platforms like Twitter and Reddit~\cite{Achimescu2020}. Its impact is amplified due to the active engagement of multiple users in the thread. Investigating these in-thread behaviors provides valuable insights into the dynamics of countering~\old{and offers guidance for the design of effective counter-misinformation strategies}.

    \item 
    \new{
    \textbf{Synergizing Professionals and Crowds to Combat Misinformation: }  
    Combining professional expertise with crowd insights offers a more comprehensive approach to tackling misinformation. 
    Professionals can guide and train the crowd, ensuring contributions adhere to fact-checking standards and best practices. 
    For instance, when creating counter-misinformation responses, crowds could learn from fact-checkers on how to craft evidence-based responses, properly cite sources, and maintain objectivity. 
    Conversely, the crowd provide diverse perspectives and identify emerging misinformation trends, aiding fact-checkers in staying informed. 
    Fostering this symbiotic relationship between professionals and crowds enhances effectiveness in countering misinformation. 
    }

    \item 
    \new{

    \textbf{Identifying and Mitigating Attack in Crowd-based Fight against Misinformation: }
The crowd-based approach to combating misinformation, while promising, faces risks of coordinated attacks and exploitation by bad actors. 
This future research direction would involve identifying and analyzing such potential attacks and developing strategies to mitigate them. 
This is common considering the anonymity and decentralized nature of crowdsourcing. Finally, bad actors can amplify their influence or spread contradictory information.
Potential strategies to mitigate these risks include robust identity verification procedures, moderation systems to filter out false information, and continuous monitoring and adaptation to evolving threats: 
(1) Build robust identity verification and reputation systems to filter out bad actors and promote accountability within the community;
(2) Implement moderation and quality control systems to remove false messages before they propagate further, which can be achieved by automated detection and filtering mechanisms;
(3) Continuously adapt and improve to stay ahead of potential threats, as attack vectors may evolve over time. 
    }

    \item
    \new{
    \textbf{Leveraging Large Language Models to Empower Crowds to Combat Misinformation: }
    Recent advancements in large language models (LLMs) present an exciting opportunity to enhance and augment the capabilities of crowds' efforts in combating misinformation. 
    Potential directions include:
    (1) Screening and annotating potential misinformation to prioritize and surface content that exhibits patterns or signals of misinformation, and flag it for further human evaluation; 
    (2) Evaluating the counter-misinformation text by synthesizing
    information from multiple sources to verify claims and identify potential inconsistencies within the misinformation text. 
    (3) Creating high-quality counter-misinformation 
    tailored to specific audiences, languages, and cultural contexts to amplify the reach and impact of the efforts. 
    However, it is crucial to acknowledge the potential risks and limitations associated with LLM integration, such as the propagation of biases, hallucinations, or the generation of plausible but factually incorrect content.  
    }

\end{itemize}

\section{Conclusions}

While crowd-based efforts to combat misinformation have increasingly attracted attention, there has yet to be a comprehensive survey paper that examines the multifaceted roles of crowds. 
Our study fills this gap by systematically categorizing the three primary roles of crowds: annotators, evaluators, and creators.
Toward this end, we present the inaugural systematic survey of 88 papers investigating crowd-based efforts, which were collected by following PRISMA guidelines.
Specifically, we presented key data statistics on misinformation, counter-misinformation, and crowd inputs found in the literature.
Additionally, we proposed a novel taxonomy that covers the diverse roles of crowds in a comprehensive way. 
Moreover, we provide detailed insights and findings extracted from the surveyed papers, offering valuable resources for effective counter-misinformation. 
By doing so, this survey helps readers grasp the latest research developments in this field and establishes the foundation for encouraging advanced crowd-assisted methodologies to combat misinformation.

\textbf{{ACKNOWLEDGMENTS}} 
This research/material is based upon work supported by NSF grant CNS-2154118. Any opinions, findings and conclusions or recommendations expressed in this material are those of the author(s) and do not necessarily reflect the position or policy of NSF and no official endorsement should be inferred.

\bibliographystyle{ACM-Reference-Format}
\bibliography{main} 

\end{document}